\documentclass[fleqn, usenatbib]{mnras}

\input{Preamble/packages}

\newcommand{\KTH}{KTH Royal Institute of Technology, Department of Physics, SE-10691 Stockholm, Sweden}
\newcommand{\OKC}{The Oskar Klein Centre for Cosmoparticle Physics, AlbaNova University Centre, SE-10691 Stockholm, Sweden}

\newcommand{\cm}[1]{{ \rm{\, cm^{#1}} }} 
\newcommand{\s}[1]{{ \rm{\, s^{#1}} }} 
\newcommand{\keV}[1]{{ \rm{\, keV^{#1}} }} 
\newcommand{\MeV}[1]{{ \rm{\, MeV^{#1}} }} 
\newcommand{\fluence}{{ \rm{\, erg \, cm^{-2}} }} 
\newcommand{\pflux}{{ \rm{\, cm^{-2} \, s^{-1}}}} 

\newcommand{\rph}[1]{r_{ \rm{ph}#1 }} 
\newcommand{\rcoll}[1]{r_{ \rm{coll}#1 }} 

\newcommand{\tvar}[1]{t_{ \rm{var}#1 }} 
\newcommand{\Fob}[1]{F_{ \rm{ob}#1 }} 
\newcommand{\Nob}[1]{N_{ \rm{ob}#1 }} 
\newcommand{\Tob}[1]{T_{ \rm{ob}#1 }} 
\newcommand{\kB}[1]{k_{ \rm{B}#1 }} 
\newcommand{\eff}[1]{\epsilon_{ \gamma#1 }}
\newcommand{\epsmax}[0]{\epsilon_{\rm{max}}}
\newcommand{\epsd}[0]{\Bar{\epsilon}_d}
\newcommand{\tu}[1]{\theta_{ u#1 }}
\newcommand{\tuk}[1]{\theta_{ u,\rm{K}#1 }}
\newcommand{\tr}[1]{\theta_{ r#1 }}
\newcommand{\tC}[1]{\theta_{ \rm{C}#1 }}

\newcommand{\lr}[1]{ \left( #1 \right) } 
\newcommand{\lrb}[1]{ \left[ #1 \right] } 

\title[Radiation-mediated shock in GRB 211211A]{Photospheric emission from GRB 211211A altered by a strong radiation-mediated shock}

\author[Wistemar, Alamaa, Ryde]{
Oscar Wistemar,$^{1, 2}$\thanks{E-mail: wistemar@kth.se}
Filip Alamaa,$^{1, 2, 3}$
Felix Ryde$^{1, 2}$
\\
$^{1}$\KTH \\
$^{2}$\OKC \\
$^{3}$IAP Institut d’Astrophysique de Paris, Centre National d’Etudes Spatiales, Sorbonne Université, UMR 7095, 98 bis Arago,\\ F-75014 Paris, France
}

\date{Accepted 2025 October 10. Received 2025 October 10; in original form 2025 October 10}

\pubyear{2025}

\begin{document}
\label{firstpage}
\pagerange{\pageref{firstpage}--\pageref{lastpage}}
\maketitle

\begin{abstract}
    Gamma-ray burst (GRB) spectra are typically non-thermal, with many including two spectral breaks suggestive of optically-thin emission. However, the emitted spectrum from a GRB photosphere, which includes prior dissipation of energy by radiation-mediated shocks (RMSs), can also produce such spectral features. Here, we analyze the bright GRB 211211A using the Kompaneets RMS Approximation (KRA). We find that the KRA can fit the time-resolved spectra well, significantly better than the traditionally used Band function in all studied time bins. The analysis of GRB 211211A reveals a jet with a typical Lorentz factor ($\Gamma \sim 300$), and a strong RMS (upstream dimensionless specific momentum, $\gamma_u \beta_u \sim 3$) occurring at a moderate optical depth ($\tau \sim 35$) in a relatively cold upstream ($\theta_u = k_{\rm B} T_u / m_e c^2 \sim 10^{-4}$). We conclude that broad GRB spectra that exhibit two breaks can also be well explained by photospheric emission. This implies that, {in such cases}, the spectral shape in the MeV-band alone is not enough to determine the emission mechanism during the prompt phase in GRBs.
\end{abstract}


\section{Introduction}\label{sec:intro}    

The initial flash of gamma-rays in a gamma-ray burst (GRB) is referred to as the prompt emission and has been studied in detail, however, the emission mechanism behind it is still not fully understood. Observed spectra of GRBs are typically broad with a broken power-law like shape spanning a few orders of magnitude in energy.

One possibility is that synchrotron radiation from optically thin parts of the jet is responsible for the observations. The synchrotron emission could come from internal collisions within the flow \citep{Narayan92, ReesMeszaros94, Kobayashi97, Daigne98}, magnetic dissipation \citep{Spruit01, Drenkhahn02, ICMART_Zhang11}, or from interaction with the immediate circumburst material \citep{Ryde22, Peer24}. Synchrotron spectra are generally broad \citep[e.g.,][]{Katz94, Tavani96} and have thus been used to fit GRB spectra \citep{Burgess11, Zhang16}. However, a limiting factor is that synchrotron spectra are too soft for a significant fraction of GRBs \citep[synchrotron line of death][]{Preece98, Acuner20}.

An alternative is that the prompt emission comes from the photosphere, which is the radius where the optical depth to a distant observer is unity. An important realization is that spectra from the photosphere are much broader than the naively expected Planck or Wien distributions, making them viable contenders to explain the observations. There are two main reasons for this. The first reason is due to geometrical broadening by emission from multiple radii and angles. Photons from the photospheric region are emitted at different radii (since the decoupling is a probabilistic process), and as the outflow expands adiabatically, the energy of the average photon decreases \citep{Peer08, Beloborodov11, Lundman13}. Photons are also emitted at different azimuthal angles (the angle from the line of sight, i.e., high latitudes) and thus experience various Lorentz boosts \citep{Abramowicz91, Peer08}. Further effects from emission at different azimuthal angles arises for structured jets \citep{Lundman13, Meng18, Hu25}, where the Lorentz factor and the emitted energy is different at different angles. Therefore, while the local, comoving emission can indeed have a Planck (or Wien) distribution, the observed spectrum is significantly broader, forming a shape that is often referred to as a non-dissipated photospheric spectrum \citep[NDP,][]{Acuner19}. 

The second reason, which is the most important one for this paper, is that any significant energy-dissipation occurring in the outflow alters the energy distribution of the photons \citep{Rees05}. Dissipation in the jet is naturally caused by shocks due to the highly variable flow \citep[e.g., ][]{Lazzati09}. Such shocks, occurring below the photosphere, are radiation-mediated shocks \citep[RMSs;][]{Levinson08}, and the shock dictates how photons gain and lose energy \citep{Ito18, Lundman18}. Since the dissipation occurs below the photosphere, the shocked outflow is still optically thick. The downstream photon field then experiences thermalization as it propagates towards the photosphere. If the dissipation occurs close enough below the photosphere, the photon field will not have to time to completely re-thermalize, and the emitted spectrum at the photosphere contains information about the dissipation. However, the deeper the dissipation occurs, the more information is lost to thermalization. 

In the literature there are many examples of GRBs with multiple breaks in their prompt spectrum \citep[e.g., ][]{Strohmayer98, Barat98}. Such spectra have been interpreted as (i) synchrotron emission in the marginally fast-cooling regime \citep{Oganesyan17, Ravasio18, Burgess20, Ghisellini20}, (ii) synchrotron emission with an additional subdominant blackbody (BB) component \citep[e.g.][]{Battelino07, Guiriec11, Iyyani13}, and finally, (iii) synchrotron emission, in which the contribution from the reverse and the forward shock both peak in the observed gamma-ray band \citep{Daigne98, Rahman24}. In addition to these possibilities, \citet{Samuelsson23} studied the observational characteristics of RMSs and found that RMS spectra also can produce spectral shapes with two breaks. 

In this paper, we study GRB 211211A, a bright burst that have two breaks in the spectrum, which has consequentially been interpreted as marginally fast-cooling synchrotron \citep{Gompertz23, Mei25} and a nonthermal component with an additional subdominant BB \citep{Peng24}. Here, we investigate if a photospheric framework with an RMS model \citep[the KRA,][]{Samuelsson22} can explain the observations. We find that the RMS model performs well, outperforming the Band function \citep{Band93} in all time-resolved spectra, thus providing a new interpretation of the observations.

The paper is organized as follows: in \S \ref{sec:GRB211211A} we introduce GRB 211211A and its characteristics, in \S \ref{sec:model} we go through RMSs and the KRA model, in \S \ref{sec:gamma} we measure the Lorentz factor, in \S \ref{sec:results_spectra} we present the result of the time-resolved spectral analysis, in \S \ref{sec:results_RMS} we study the RMS evolution, in \S \ref{sec:discussion} we discuss our results, and finally in \S \ref{sec:conclusion} we conclude.

\section{Prompt emission of GRB 211211A}\label{sec:GRB211211A}

GRB 211211A was detected by the {\it Fermi} Gamma-ray Burst Monitor \citep[GBM;][]{Fermi_GBM} and the {\it Swift} Burst Alert Telescope \citep[BAT;][]{Swift_BAT} at $T_0 =$ 13:09:59 UTC \citep{Fermi21A, Swift21A}. The GBM fluence is $5.4\times10^{-4}\fluence$ (10$-$1000 keV; 1.3$-$54.0 s) and the $T_{90}$ is 34.3 s (50$-$300 keV) \citep{Fermi21B}. The BAT fluence is $1.5\times10^{-4}\fluence$ (15$-$150 keV; 0$-$142 s) and the $T_{90}$ is 51.37 s (15$-$350 keV) \citep{Swift21B}. It has a measured redshift of $z = 0.076$ \citep[][]{redshift21}. It is an example of a burst with a broad prompt gamma-ray spectrum, which has two spectral breaks \citep{Gompertz23, Peng24}. 

GRB 211211A has been widely studied by other authors due to its brightness and contradictory origin. Its duration is that of a long GRB ($T_{90} > 2 \s{}$), however, the discovery of an associated kilonova \citep{Rastinejad22, Troja22} strongly indicates that the burst originated from a compact object merger (NS-NS or NS-BH), usually expected to form short GRBs ($T_{90} < 2 \s{}$). However, an alternative has been raised by \citet{Waxman25} where the excess emission instead is explained as emission from dust heating in connection with a supernova.

\begin{figure}
    \includegraphics[width = \columnwidth]{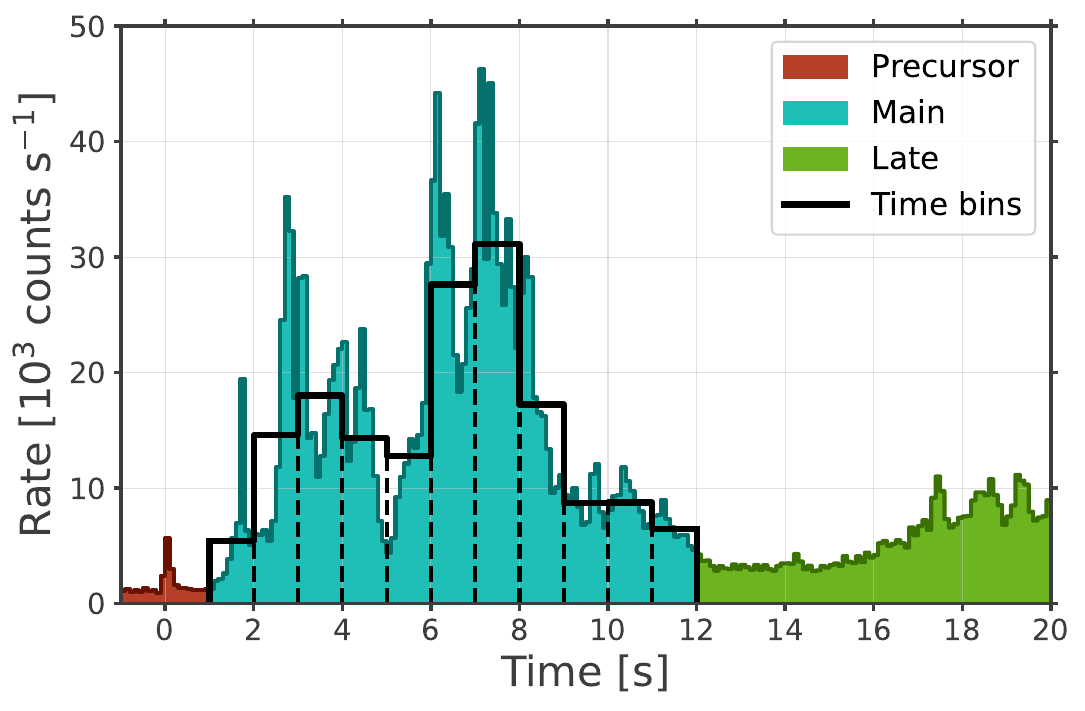}
    \caption{The Fermi GBM light curve of GRB 211211A from $T_0 - 1$ s to $T_0 + 20$ s. The red, blue, and green colors represent the precursor, the main emission, and the late/extended emission, respectively. The black histogram shows the binned light curve used in the analysis.}
    \label{fig:light_curve_full}
\end{figure}

The light curve observed by the {\it Fermi} GBM detector is shown in Figure \ref{fig:light_curve_full}. It is typically divided into three parts \citep[e.g.][]{Veres23, Xiao24}: The pre-cursor that caused the trigger (0$-$1 s), the main emission (1$-$12 s), and the late/extended emission (EE) that continues until the GBM detection falls below the background level (12$-$ $\sim$ 60 s). In this paper we study the main emission, which we divide into eleven one second long time bins following \citet{Veres23}.

In \S \ref{subsec:measure_gamma}, we employ the light curve variability time to estimate the bulk outflow Lorentz factor. The variability time in the light curve of GRB 211211A has been measured with several different assumptions and methods, arriving at different results \citep{Veres23, Yang22, Xiao24, Troja22, Camisasca23, Peng24}. Here, we use the time resolved variability time from \citet{Veres23}, which motivates the choice of time bins.

\section{Modelling radiation-mediated shocks}\label{sec:model}

\subsection{Subphotospheric radiation-mediated shocks}\label{subsec:RMS}

GRB prompt emission is commonly modeled by internal shocks occurring in the jet outflow \citep{ReesMeszaros94}. When these shocks occur well below the photosphere in GRBs, they are mediated by radiation \citep[for a review on RMSs, see e.g.][]{Levinson20}. Such shocks are expected to be photon rich \citep{Bromberg11}, meaning that the photons generated in the shock and the immediate downstream are negligible compared to the number of advected upstream photons.

An RMS propagating into a thermal upstream is characterized by three parameters \citep[see for instance][]{Lundman18}. (i) The upstream velocity\footnote{Subscripts $u$ and $d$ are used to indicate quantities measured far upstream and downstream of the shock, respectively.}, $\beta_u = v_u / c$, where $v_u$ is the fluid velocity relative to the shock frame. Together with the upstream Lorentz factor, $\gamma_u = (1-\beta_u^2)^{-1/2}$, it gives the dimensionless specific momentum, $u_u = \gamma_u \beta_u$. (ii) The photon-to-proton ratio, $\zeta = n_\gamma / n_p$, where $n_\gamma$ ($n_p$) is the comoving number density of photons (protons). This ratio is preserved across the shock transition when the shock is photon rich. (iii) The upstream comoving temperature, $\tu{} = \kB{} T_u / m_e c^2$, where $T$ is the temperature in Kelvin, $\kB{}$ is the Boltzmann constant, $m_e$ is the electron mass, and $c$ is the speed of light. Throughout the rest of the paper, $\theta$ represents comoving frame dimensionless temperatures in units of the electron rest mass energy. These three parameters, together with the optical depth where the shock is initiated, $\tau$, determine the photospheric spectrum.

When photons traverse an RMS, they experience bulk Comptonization. They scatter in the speed gradient across the shock, which lead to a Fermi-like process where the photons gain energy on average with each scattering. This process produces a power-law photon spectrum in the immediate downstream. Nonrelativistic and mildly relativistic RMS spectra typically have two breaks, a high-energy and a low-energy one. In the immediate downstream, the high-energy break is given by $\epsmax = \langle \Delta \epsilon / \epsilon \rangle$, which is the maximum energy a photon can get in the shock before it is significantly effected by electron recoil. Here, $\langle \Delta \epsilon / \epsilon \rangle $ is the average fractional energy gain in a scattering in the shock, which is found to be proportional to $u_u^2$ \citep[][]{Blandford81, Samuelsson22}, and $\epsilon$ is the photon energy in units of the electron rest mass energy.

\begin{figure}
    \centering
    \includegraphics[width = \columnwidth]{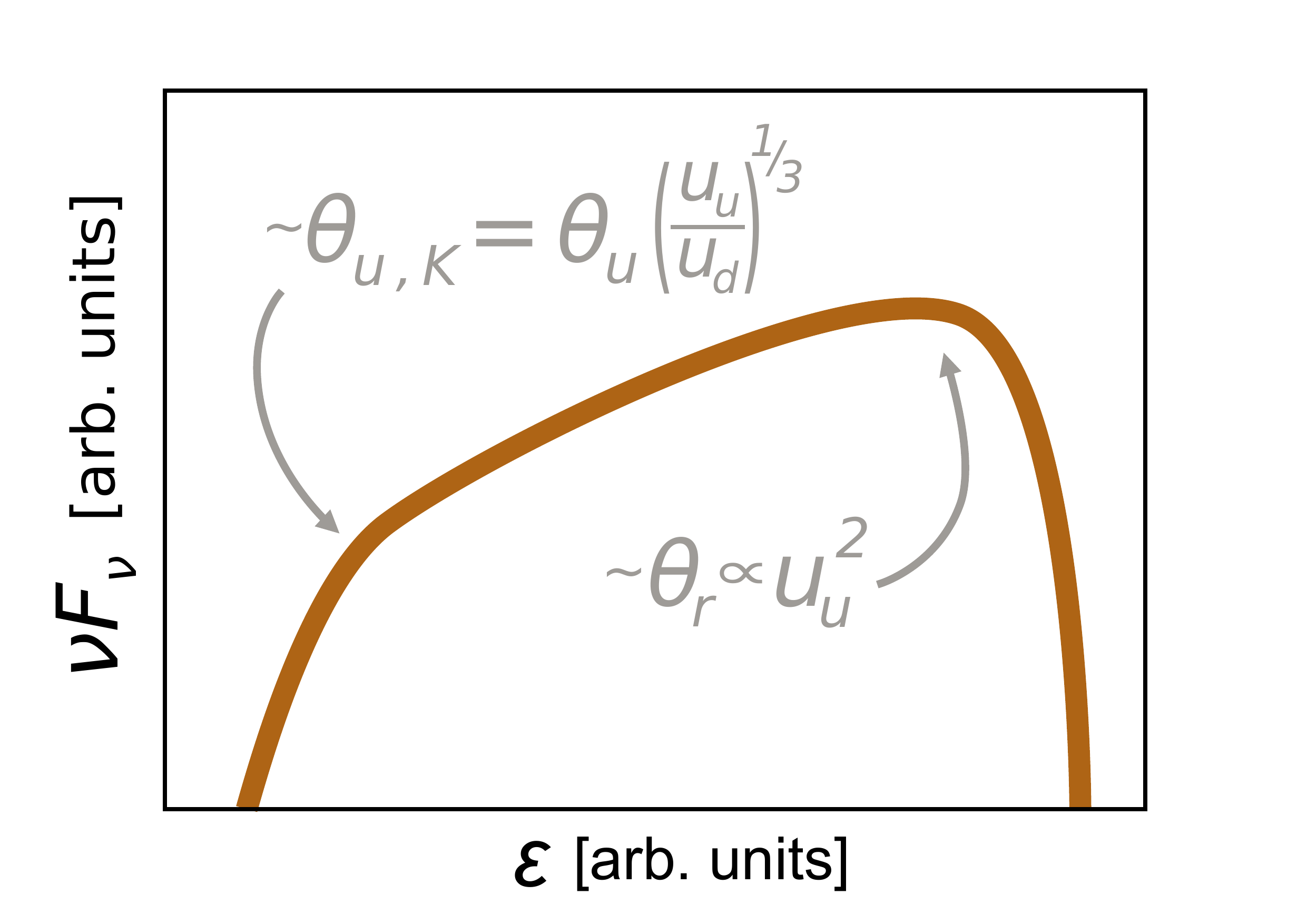}
    \caption{A schematic of a typical immediate downstream spectrum in $\nu F_\nu$ representation. The two breaks are shown with their respective parameter dependencies. Here, $\tu{}$ is the upstream temperature and $u_u$ is the dimensionless specific momentum in the RMS model.}
    \label{fig:schematic}
\end{figure}

The low-energy break in the immediate downstream is at $\tu{} (u_u / u_d)^{1/3}$. The factor $(u_u / u_d)^{1/3}$ comes from the adiabatic compression across the shock, which increases the temperature \citep{Blandford81}. These photons only gain energy from the compression and not via bulk Comptonization in the shock. An immediate downstream spectrum with the two breaks, and their parameter dependencies, is visualized in the schematic in Figure \ref{fig:schematic}.

These breaks are clearly distinguishable in the immediate downstream spectrum. However, the prominence of these breaks in the emitted spectrum depends on the optical depth, $\tau$, which affects the overall thermalization of the spectrum. Furthermore, the bulk outflow Lorentz factor $\Gamma$ determines if these breaks are within the observed energy range of the detector.

\subsection{The Kompaneets RMS Approximation}\label{subsec:KRA}

Full hydrodynamic simulations, coupled with Monte Carlo photons, of nonrelativistic and mildly relativistic RMSs produce typical GRB spectra with a smooth broad power law between the aforementioned two breaks \citep{Ito18, Lundman18}. However, these simulations of RMSs are computationally expensive and therefore the Kompaneets RMS approximation (KRA) was introduced in \citet{Samuelsson22}.

The KRA is an analogous model to an RMS in a GRB outflow, based on the similarity between the energy dissipation process in RMSs and thermal Comptonization. The KRA splits the continuous RMS region into three discrete zones: the upstream, the RMS, and the downstream zone. In each zone, photons interact with the plasma via thermal Comptonization and the zones are evolved by solving the Kompaneets equation. Energy dissipation in the RMS zone is achieved by prescribing the electrons with a high effective temperature, $\tr{}$. The zones are connected via source terms and photons flow from the upstream, via the RMS zone, to the downstream. The upstream is assumed to be in a thermal Wien distribution, which implies that if previous dissipation has occurred the photons have had time to reestablish a kinetic equilibrium with the plasma.

In \citet{Samuelsson22} it was shown that the KRA is accurate up to mildly relativistic shocks. The faster computational speed of the KRA allows a table model to be created, which can be used when fitting data. Since the KRA is an analogous model it has a different set of parameters and KRA spectra are characterized by the following three parameters. (i) The effective temperature in the RMS zone, $\tr{}$. (ii) The Compton $y$-parameter in the RMS zone, $y_r$, which is the product of the average energy gain and mean number of scatterings in the RMS zone. (iii) The upstream temperature, $\tuk{}$, which is slightly larger than $\tu{}$, due to the adiabatic compression (see Equation \eqref{eq:theta_u}). Finally, similarly to the RMS picture, the photospheric spectrum is also dependent on the optical depth, $\tau$. 

When fitting, there are two additional parameters from the ones mentioned above, the normalization $K$ and the parameter $S$ that shifts the spectrum along the energy axis \citep{3ML_Vianello15}. In our setup, the $S$-parameter corresponds to the bulk Lorentz factor of the outflow, $\Gamma$.

The similarity between the RMS and the KRA spectra implies that the KRA spectra also have two breaks, a high-energy and a low-energy one. For KRA, the high-energy break is determined by $\tr{}$, since in thermal Comptonization the electron temperature is related to the average fractional energy gain as $4\theta = \langle \Delta \epsilon / \epsilon \rangle$. Therefore, the maximum energy is $\epsmax = 4\tr{}$. The low-energy break in the immediate downstream corresponds to $\tuk{}$, i.e., the compressed upstream temperature (see Figure \ref{fig:schematic} for a schematic spectrum).

\citet{Samuelsson22, Samuelsson23, Wistemar23} found that the KRA model can fit observed GRB spectra very well, however, \citet{Samuelsson22} found that there is a strong degeneracy when fitting observed spectra. They identified that comoving spectra at different energies are identical in shape if the parameter combinations $\tr{} \tau$, $\tr{}/ \tuk{}$, and $y_r$ remain constant. Thus, with an unconstrained $\Gamma$, these can become identical observed spectra. Due to this degeneracy, only the spectral shape is known from a fit and not the absolute position in the comoving frame. However, once the Lorentz factor is known, this degeneracy is broken and one can retrieve the comoving absolute position from the observed spectrum, thus resolving $\tr{}$ and $\tuk{}$.

\subsection{Conversion between RMS and KRA parameters}\label{subsec:convert}

The conversion between the three RMS and the three KRA parameters is described in detail in \citet{Samuelsson22}. Here, we give a brief summary. As mentioned before, the KRA does not account for the shock compression, and therefore the temperature difference between $\tuk{}$ and $\tu{}$ is the compression factor \citep[Equation (A6) in][]{Samuelsson22}

\begin{equation}\label{eq:theta_u}
    \tuk{} = \tu{} \left(\frac{u_u}{u_d}\right)^{1/3}.
\end{equation}

The RMS parameter $u_u$ roughly determines the strength of the shock and is connected to $\tr{}$, as it determines the ``strength'' of the RMS zone in the KRA model. This relation is given as \citep[Equation (5) in][]{Samuelsson22}

\begin{equation}\label{eq:theta_r}
    4\tr{} = \frac{(u_u)^2 \ln(\Bar{\epsilon}_d/\Bar{\epsilon}_u)}{\xi}.
\end{equation}

\noindent Here, $\Bar{\epsilon}$ is the average comoving photon energy and $\xi$ is a constant. Empirically, it was found that $\xi = 55$ gave good agreement between the KRA and full-scale radiation hydrodynamic simulations \citep{Lundman18} over a wide range of initial parameters \citep{Samuelsson22}. 

The third relation is found by requiring that the average photon energy in the immediate downstream, ${\bar \epsilon}_d$, is the same in both models. Using Equations \eqref{eq:theta_u} and \eqref{eq:theta_r}, together with the shock jump conditions, gives a one-to-one correspondence between the KRA-parameter $y_r$ and the RMS parameter $\zeta$. 

For multiple of these conversions, we need a way of estimating ${\bar \epsilon}_d$ as a function of the KRA parameters. A quick and accurate empirical method is presented in \citet{Bagi25}, which will be used here.

\section{Measuring the bulk outflow Lorentz factor}
\label{sec:gamma}

The KRA model requires knowledge of the bulk outflow Lorentz factor in order to relate observed spectra to comoving spectra (see \S \ref{subsec:KRA}). In this Section, we apply the method derived in \citet{Wistemar25} to measure the Lorentz factor. In \S \ref{subsec:measure_gamma}, we briefly reiterate the method and apply it to the KRA model and in \S \ref{subsec:observed_properties}, we use the method to measure the Lorentz factor of GRB 211211A.

\subsection{\texorpdfstring{Measuring $\Gamma$ from observed properties}{Measuring the Lorentz factor from observed properties}}\label{subsec:measure_gamma}

\citet{Wistemar25} introduced a generalized method to measure $\Gamma$ from photospheric, prompt emission, which can also be used for highly dissipative flows that may lead to nonthermal spectral shapes. They presented multiple ways of measuring $\Gamma$ depending on what observations are available for a particular GRB. Here, we use their Equation $(9)$:

\begin{equation}\label{eq:gamma_tvar}
    \Gamma = \lr{\frac{2.7 k_B}{2c^2 \sigma}}^{1/2} \frac{d_L}{(1 + z)} \frac{\Nob{}^{1/2}}{\Tob{}^{3/2}\tvar{}},
\end{equation}

\noindent where $z$ is the redshift, $d_L$ is the luminosity distance, $\Nob{}$ is the photon flux in the observer frame, $\sigma$ is the Stefan-Boltzmann constant, and $\tvar{}$ is the characteristic variability time of the outflow, assumed to equal $\rph{} (1 + z) / 2 c \Gamma^2$. The observed temperature, $\Tob{}$, corresponds to $\tuk{}$ after having accounted for adiabatic cooling and the Doppler boost \citep[see][and Equation \eqref{eq:Tob} below]{Wistemar25}. In GRBs with an additional break in X-rays, like GRB 211211A, $\Tob{}$ is related to this break. For the luminosity distance we use a cosmology with values from \citet{Planck20}.

In order to determine $\Gamma$, we thus need four parameters from the observations: $z$, $\Nob{}$, $\Tob{}$, and $\tvar{}$. Since spectra from the KRA model fit observed data very well, it is straightforward to integrate a fitted KRA spectrum to obtain $\Nob{}$. The observed temperature corresponds to a feature of the spectrum. It can be found from the fitted KRA parameters by (see Equation 3 in \citet{Wistemar25})

\begin{equation}\label{eq:Tob}
    \Tob{} = \frac{2 m_e c^2 \tuk{} S}{(1 + z) k_B \tau^{2/3}},
\end{equation}

\noindent where $\tuk{}$ is used instead of their comoving temperature at the shock $T^\prime$, and the scale parameter $S$ for the Lorentz factor. Due to the degeneracy mentioned in \S \ref{subsec:KRA}, there are several different combinations of the fitted KRA parameters with identical observed spectra. However, these different combinations still result in similar values for $\Nob{}$ and $\Tob{}$. This is expected since $\Nob{}$ and $\Tob{}$ are properties of the observed spectrum, which all allowed parameter combinations must satisfy. Finally, the variability time can be estimated from the light curve and the redshift can be obtained via photometry or spectroscopy.

\subsection{\texorpdfstring{Evolution of $\Gamma$ in GRB 211211A}{Evolution of the Lorentz factor in GRB 211211A}}\label{subsec:observed_properties}

To determine $\Nob{}$ and $\Tob{}$ for the time-resolved spectra of GRB 211211A, we perform fits with the KRA model. We focus on the main emission episode of GRB 211211A, starting at $T_0 + 1$ s until $T_0 + 12$ s, with the previously mentioned one second long time bins (see Figure \ref{fig:light_curve_full}). The data is taken from the GBM detectors on {\it Fermi}, more specifically, the detectors NaI2, NaI10, and BGO0. The Multi-Mission Maximum Likelihood framework \citep[3ML;][]{3ML_Vianello15} is used for the data analysis with its data processing tools and Bayesian inference setup. We use the nested sampler \verb|UltraNest| \citep[]{ultranest} for the inference, with log-uniform priors for $K \sim$ log $\mathcal{U} (10^{-4}-10^{4})$, $S \sim$ log $\mathcal{U} (10^{1}-10^{3})$, and $\tr{} \sim$ log $\mathcal{U} (0.01-0.25)$, along with a uniform prior on $y_r \sim \mathcal{U} (0.1-6)$. For technical reasons, we use a uniform prior on the product $\tau \tr{} \sim \mathcal{U} (1-50)$ and a log-uniform prior on the ratio $\tr{} / \tuk{} \sim$ log $\mathcal{U} (30-10^{4}$), see \citet{Samuelsson22} for an explanation of the parameter combinations.

\begin{figure}
    \centering
    \includegraphics[width = \columnwidth]{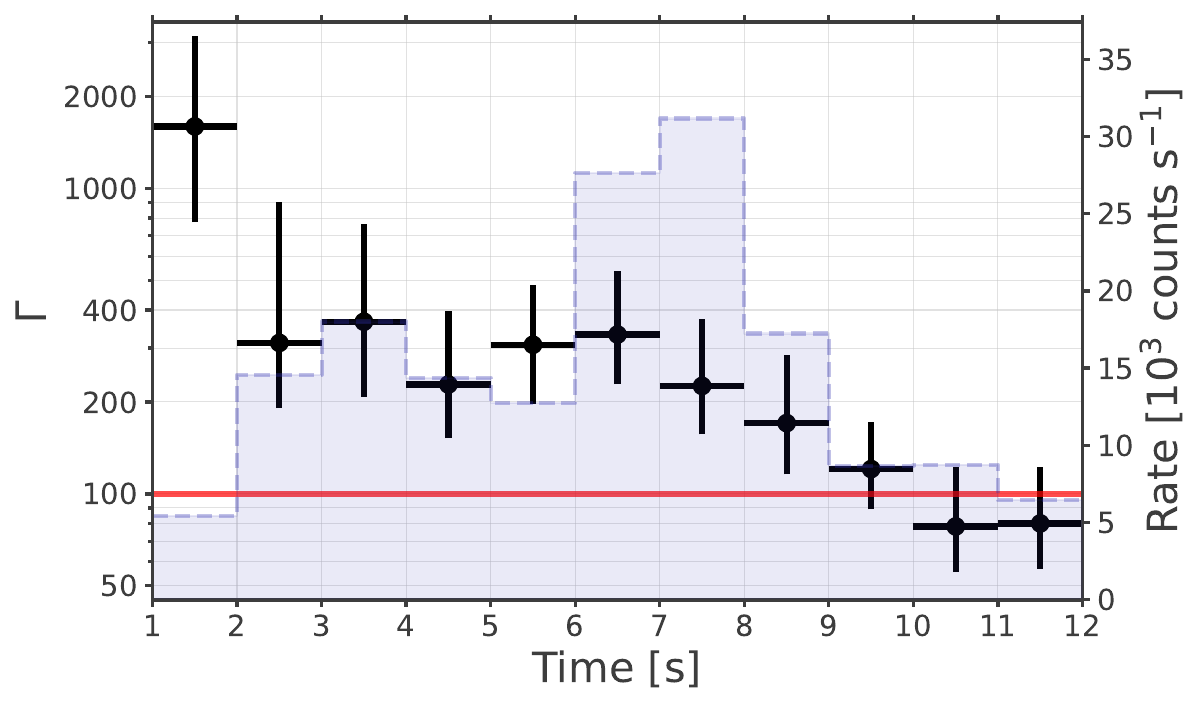}
    \caption{The measured values of $\Gamma$, using the method outlined in \S \ref{sec:gamma}. The black dots show the best-fit values with the 68\% confidence interval displayed by the error bars. The best fit values are used for the analysis in \S \ref{sec:results_spectra}. The red line shows the constant value of $\Gamma = 100$ used for comparison in \S \ref{sec:results_RMS}. The light blue histogram is the binned light curve, with values on the y-axis to the right.}
    \label{fig:gammas}
\end{figure}

The Bayesian inference with the KRA model results in well determined values for $\Nob{}$ and $\Tob{}$. We use the time-resolved variability time from \citet{Veres23} as $\tvar{}$, (see \S \ref{sec:GRB211211A}), and the measured redshift of $z = 0.076$ from \citet{redshift21}. Finally, the Lorentz factor in each time bin is obtained from Equation \eqref{eq:gamma_tvar}. The result and the values found are shown in Table \ref{tab:gamma} and Figure \ref{fig:gammas}. The analysis shows that the Lorentz factor evolves with time over the burst, decreasing from $\gtrsim 1000$ to $\lesssim 100$. For the most luminous part of the GRB (2-9 s) the Lorentz factor only varies slowly and remains at around $150-350$. The errors of the measurement are around $30\%$, which is found by propagating the errors from $\Tob{}$, $\Nob{}$, and $\tvar{}$ in Equation \eqref{eq:gamma_tvar}.

With the Lorentz factor determined, it is straightforward to calculate the photospheric radius using the variability time as 

\begin{equation}\label{eq:r_ph}
    \rph{} = \frac{2 c \Gamma^2 \tvar{}}{1 + z}.
\end{equation}

\noindent For the most luminous part of the burst (2-9 s), the photospheric radius is $\rph{} \sim 2 \times 10^{13} \cm{}$. For the first time bin, the large Lorentz factor results in a larger emission radius at $\sim 4 \times 10^{14}$ cm.

The Lorentz factor of GRB 211211A has been previously estimated in the literature \citep{Rastinejad22, Mei22, Veres23, Peng24}. The estimates vary from 70 \citep{Rastinejad22} to $\geq 1000$ \citep{Mei22, Veres23}. Our measurements lie within this interval, similar to the value found in \citet{Peng24} and compatible within the $1\sigma$ error to the value found in \citet{Mei22}. However, we note that the methods used in these works are different.

\begin{table*}
	\centering
	\caption{Determination of the Lorentz factor, $\Gamma$, as explained in detail in \S \ref{sec:gamma}. The values used to determine $\Gamma$ are also displayed. The errors for $\Nob{}$ and $\Tob{}$ are from the posterior, $\tvar{}$ and its errors are from \citet{Veres23}, and the errors for $\Gamma$ are propagated from Equation \eqref{eq:gamma_tvar}. All errors represent the 68\% confidence interval and the redshift, $z = 0.076$, is taken from \citet{redshift21}.}
	\label{tab:gamma}
    {\setlength{\extrarowheight}{5pt}
	\begin{tabular}{ccccc} 
		\hline
		Time [s] & $\Nob{} \lrb{\pflux{}}$ & $\kB{}\Tob{} \lrb{\keV{}}$ & $\tvar{} \lrb{10^{-3}\s{}}$ & $\Gamma$ \\[4pt]
		\hline
        1-2   & $60^{+3}_{-2}$  & $2.1^{+1.1}_{-0.6}$  & $2.6^{+0.9}_{-0.9}$  & $1696^{+1550}_{-820}$ \\
        2-3   & $170^{+3}_{-2}$ & $8.6^{+0.5}_{-4.5}$  & $3.0^{+1.5}_{-1.5}$  & $312^{+591}_{-121}$ \\
        3-4   & $210^{+3}_{-2}$ & $6.4^{+1.6}_{-1.4}$  & $4.1^{+2.3}_{-2.3}$  & $366^{+398}_{-158}$ \\
        4-5   & $168^{+3}_{-2}$ & $6.5^{+0.8}_{-1.4}$  & $5.7^{+2.0}_{-2.0}$  & $228^{+170}_{-76}$ \\
        5-6   & $158^{+3}_{-3}$ & $3.9^{+1.2}_{-0.8}$  & $8.4^{+2.1}_{-2.1}$  & $308^{+177}_{-111}$ \\
        6-7   & $333^{+3}_{-3}$ & $6.9^{+1.3}_{-0.7}$  & $4.6^{+1.7}_{-1.7}$  & $332^{+204}_{-104}$ \\
        7-8   & $371^{+2}_{-2}$ & $10.4^{+1.0}_{-1.8}$ & $4.2^{+1.5}_{-1.5}$  & $226^{+148}_{-69}$ \\
        8-9   & $206^{+3}_{-2}$ & $6.1^{+0.8}_{-1.1}$  & $9.3^{+3.3}_{-3.3}$  & $171^{+114}_{-55}$ \\
        9-10  & $108^{+3}_{-3}$ & $4.1^{+0.6}_{-0.6}$  & $16^{+3.8}_{-3.8}$ & $121^{+51}_{-32}$ \\
        10-11 & $105^{+2}_{-2}$ & $4.9^{+0.5}_{-0.5}$  & $19^{+6.6}_{-6.6}$ & $78^{+44}_{-23}$ \\
        11-12 & $79^{+3}_{-2}$  & $4.2^{+0.6}_{-0.6}$  & $21^{+6.1}_{-6.1}$ & $80^{+42}_{-23}$ \\[4pt]
		\hline
	\end{tabular}}
\end{table*}

\section{\texorpdfstring{Spectral analysis with measured $\Gamma$}{Spectral analysis with measured Lorentz factor}}\label{sec:results_spectra}

With the Lorentz factor determined, we can now relate the observed spectra to the corresponding comoving spectra. To do this, we fix the value of $S$ to that of the measured Lorentz factor in each bin, as reported in Table \ref{tab:gamma}. This allows us to break the degeneracy discussed in \S \ref{subsec:KRA}. Thus, the free parameters are $K$, $\tau$, $\tuk{}$, $y_r$, and $\tr{}$.

Similarly to \S \ref{subsec:observed_properties}, we use Bayesian inference and the \verb|UltraNest| sampler \citep{ultranest} to fit the data in each time bin, as well as the same priors as in \S \ref{subsec:observed_properties}. Performing the inference yields the median and $68\%$ confidence interval values reported in Table \ref{tab:fit}. Also reported in Table \ref{tab:fit} are the median and the $68\%$ confidence interval values of the RMS parameters, obtained by converting the full posterior distribution using the method outlined in \S \ref{subsec:convert}. The evolution of the RMS parameters is discussed in \S \ref{sec:results_RMS}.

\begin{table*}
	\centering
	\caption{Spectral evolution of the sub-photospheric shock in GRB 211211A. The $\Gamma$ values from Table \ref{tab:gamma} have been used in the inference. The KRA parameter values and errors from the posterior are given, and they are converted into the RMS parameters, following \S \ref{subsec:convert}.  All errors represent the $68\%$ confidence interval.}
	\label{tab:fit}
    {\setlength{\extrarowheight}{5pt}
	\begin{tabular}{ccccccccc} 
		\hline
		Time [s] & $K$ & $\tau$ & $\tuk{} \lrb{10^{-4}}$ & $y_r$ & $\tr{}$ & $\tu{} \lrb{10^{-4}}$ & $u_u$ & $\zeta \lrb{10^5}$ \\[4pt]
		\hline
        1-2   & $1.9^{+0.5}_{-0.4}$ & $125^{+16}_{-16}$ & $0.3^{+0.1}_{-0.1}$ & $1.53^{+0.08}_{-0.07}$ & $0.18^{+0.05}_{-0.06}$ & $0.12^{+0.03}_{-0.02}$ & $3.0^{+0.4}_{-0.5}$ & $11.0^{+2.1}_{-2.6}$ \\
        2-3   & $3.7^{+0.4}_{-0.9}$ & $18^{+12}_{-6}$   & $1.9^{+0.2}_{-0.4}$ & $1.34^{+0.11}_{-0.11}$ & $0.08^{+0.08}_{-0.01}$ & $0.85^{+0.09}_{-0.18}$ & $2.4^{+0.9}_{-0.1}$ & $4.1^{+2.7}_{-0.5}$ \\
        3-4   & $4.2^{+1.0}_{-0.5}$ & $54^{+5}_{-15}$   & $2.4^{+0.2}_{-0.2}$ & $1.33^{+0.05}_{-0.04}$ & $0.12^{+0.10}_{-0.06}$ & $1.02^{+0.14}_{-0.13}$ & $2.8^{+0.9}_{-0.7}$ & $4.1^{+1.6}_{-1.0}$ \\
        4-5   & $2.8^{+0.7}_{-0.5}$ & $31^{+10}_{-10}$  & $2.8^{+0.2}_{-0.3}$ & $1.25^{+0.05}_{-0.06}$ & $0.09^{+0.04}_{-0.01}$ & $1.24^{+0.09}_{-0.15}$ & $2.6^{+0.5}_{-0.2}$ & $4.2^{+1.2}_{-0.6}$ \\
        5-6   & $2.4^{+0.9}_{-0.4}$ & $45^{+8}_{-12}$   & $1.7^{+0.1}_{-0.2}$ & $1.25^{+0.05}_{-0.04}$ & $0.13^{+0.08}_{-0.05}$ & $0.73^{+0.08}_{-0.09}$ & $3.0^{+0.7}_{-0.6}$ & $6.6^{+1.6}_{-1.3}$ \\
        6-7   & $3.2^{+0.3}_{-0.2}$ & $39^{+2}_{-4}$    & $2.6^{+0.2}_{-0.1}$ & $1.37^{+0.02}_{-0.02}$ & $0.18^{+0.05}_{-0.06}$ & $1.08^{+0.13}_{-0.07}$ & $3.4^{+0.4}_{-0.5}$ & $4.1^{+0.5}_{-0.7}$ \\
        7-8   & $3.7^{+0.4}_{-0.5}$ & $26^{+6}_{-4}$    & $4.2^{+0.1}_{-0.2}$ & $1.41^{+0.03}_{-0.03}$ & $0.10^{+0.04}_{-0.01}$ & $1.86^{+0.05}_{-0.12}$ & $2.6^{+0.4}_{-0.1}$ & $2.4^{+0.3}_{-0.1}$ \\
        8-9   & $1.3^{+0.2}_{-0.2}$ & $25^{+3}_{-4}$    & $3.2^{+0.1}_{-0.2}$ & $1.38^{+0.03}_{-0.03}$ & $0.15^{+0.05}_{-0.03}$ & $1.37^{+0.07}_{-0.11}$ & $3.1^{+0.4}_{-0.3}$ & $3.4^{+0.3}_{-0.2}$ \\
        9-10  & $1.5^{+0.3}_{-0.2}$ & $23^{+6}_{-5}$    & $3.0^{+0.2}_{-0.3}$ & $1.01^{+0.04}_{-0.04}$ & $0.20^{+0.03}_{-0.04}$ & $1.21^{+0.11}_{-0.12}$ & $4.0^{+0.3}_{-0.4}$ & $11.0^{+2.4}_{-2.0}$ \\
        10-11 & $1.2^{+0.3}_{-0.2}$ & $14^{+3}_{-2}$    & $3.8^{+0.3}_{-0.4}$ & $0.97^{+0.03}_{-0.03}$ & $0.23^{+0.01}_{-0.02}$ & $1.54^{+0.13}_{-0.15}$ & $4.4^{+0.1}_{-0.2}$ & $11.2^{+1.7}_{-1.5}$ \\
        11-12 & $1.6^{+0.4}_{-0.3}$ & $17^{+6}_{-5}$    & $3.6^{+0.5}_{-0.5}$ & $0.93^{+0.04}_{-0.04}$ & $0.22^{+0.02}_{-0.04}$ & $1.46^{+0.19}_{-0.21}$ & $4.3^{+0.2}_{-0.4}$ & $13.6^{+3.9}_{-3.0}$ \\[4pt]
		\hline
	\end{tabular}}
\end{table*}

\subsection{Evolution of the spectral shape}\label{subsec:spectral_shape}

\begin{figure}
    \centering
    \includegraphics[width = \columnwidth]{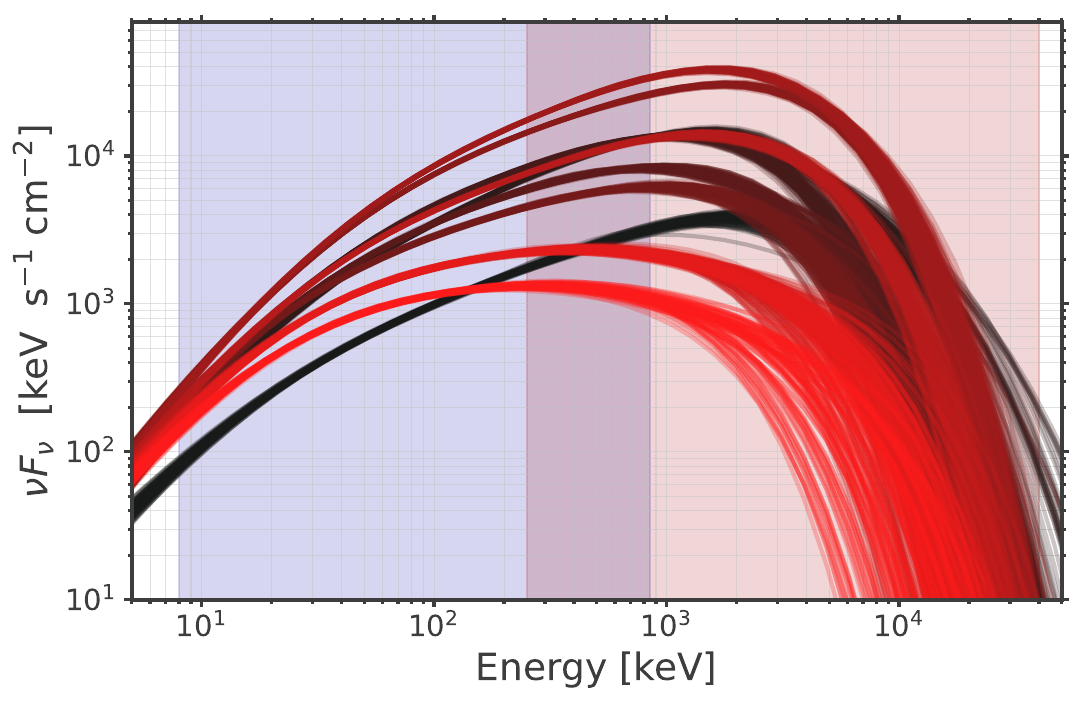}
    \caption{Time evolution of the spectra from the Bayesian inference. The darkest red is the first time bin at 1-2 s and the color brightens for each bin until the brightest red for the last bin at 11-12 s. Each color shows a set of $\sim 100$ spectra from the posterior distributions. The background shades show the energy ranges of {\it Fermi} GBM NaI (purple shade) and BGO (pink shade) detectors.}
    \label{fig:spectral_evolution}
\end{figure}

The time-evolution of the spectral shape is shown in Figure \ref{fig:spectral_evolution}, where each color shows a set of $\sim 100$ spectra from the posterior distribution for each time bin. As seen by the low spread in the posterior distribution spectra, the spectral shape is well determined from a few keV up to a few MeV. However, above $\sim 5~\MeV{}$ the shape is less well determined, mostly due to the decreasing effective area of the GBM and the lower photon flux. For the first eight time bins (1-9 s), the spectral shape is similar: they each exhibit two clear breaks with a positive (in $\nu F_\nu$) slope between the breaks. For the last three time bins (9-12 s) the shape is notably different with a softer slope. These bins are discussed further in \S \ref{subsec:relativistic} \& \S \ref{subsec:last_bins}. Note that the third and second to last sets (9-11 s) are overlapping, which is why it appears to be only ten sets of spectra in the figure.

\subsection{Time-resolved spectrum at the peak}\label{subsec:time_resolved}

\begin{figure*}
    \centering
    \includegraphics[width = 0.9\textwidth]{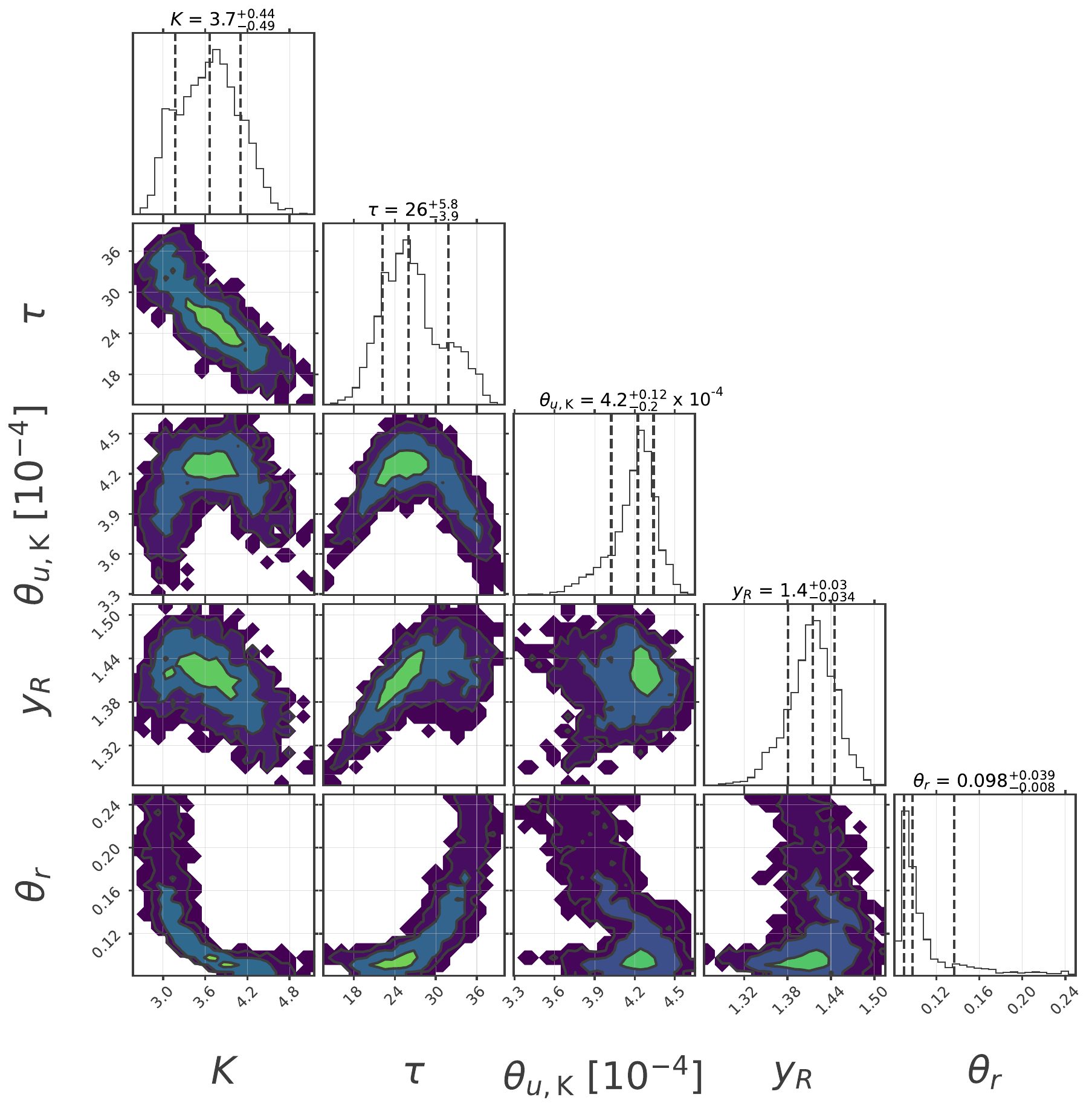}
    \caption{Corner plot of the fitted KRA parameters at 7-8 s. The center dotted line is the median value and the left and right represent the 68\% confidence interval.}
    \label{fig:corner_KRA}
\end{figure*}

\begin{figure}
    \centering
    \includegraphics[width = \columnwidth]{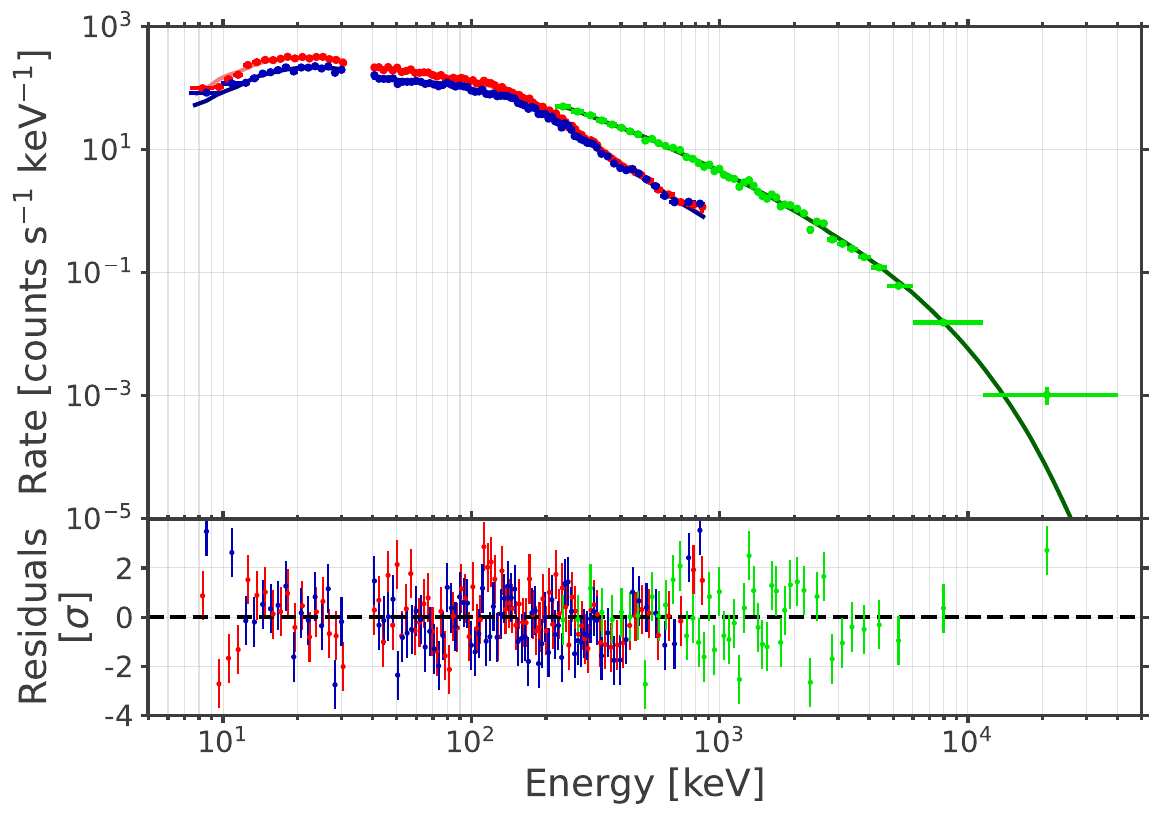}
    \includegraphics[width = \columnwidth]{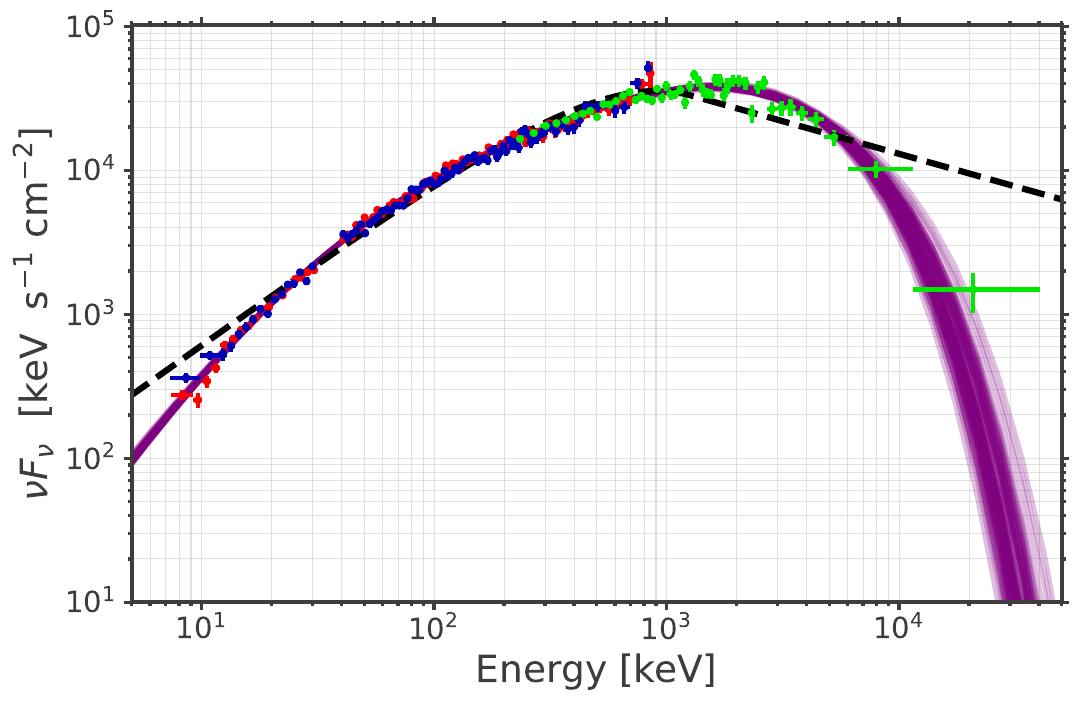}
    \caption{Spectra from the time bin at 7-8 s. The blue (NaI2), red (NaI10), and green (BGO0) data points are from the GBM onboard {\it Fermi}. \textit{Top:} Count spectrum with residuals where the lines are the fit for each detectors data. \textit{Bottom:} $\nu F_\nu$ spectrum where the purple solid lines show 1500 spectra from the posterior distribution, and the black dashed line is the best Band function fit. Note that due to the non-linearity of the GRB response matrix, data points in $\nu F_\nu$ spectra depend on the model, and thus the data points are not accurate when comparing models.}
    \label{fig:count_nuFnu}
\end{figure}

In this subsection, we focus on the peak spectrum ($7-8$ s), showing the results from the statistical inference. Figure \ref{fig:corner_KRA} shows the corner plot from the Bayesian inference with the posterior distribution of the KRA parameters. The center dashed line in the histograms is the median value while the left and right dashed lines define the 68\% confidence interval. Figure \ref{fig:count_nuFnu} shows the count spectrum with residuals (top) and the $\nu F_\nu$ spectrum (bottom) for the same time interval. The blue and red data points are from the NaI detectors (NaI2 and NaI10, respectively), while the green data points are from the BGO detector (BGO0). The random scatter in the residuals around the best fit model indicates that the fit is good. All other time bins have similarly good fits. In the $\nu F_\nu$ spectrum, the solid purple lines are spectra from the posterior distribution. The black dashed line is the best fit Band function spectrum \citep{Band93} in the same time bin, which has a low-energy index $\alpha = -0.84 \pm 0.01$, high-energy index $\beta = -2.45 \pm 0.04$, and peak energy $E_p = 911 \pm 28 \keV{}$.

\subsection{Comparison to the Band function}\label{subsec:model_comparison}

\begin{figure}
    \centering
    \includegraphics[width = \columnwidth]{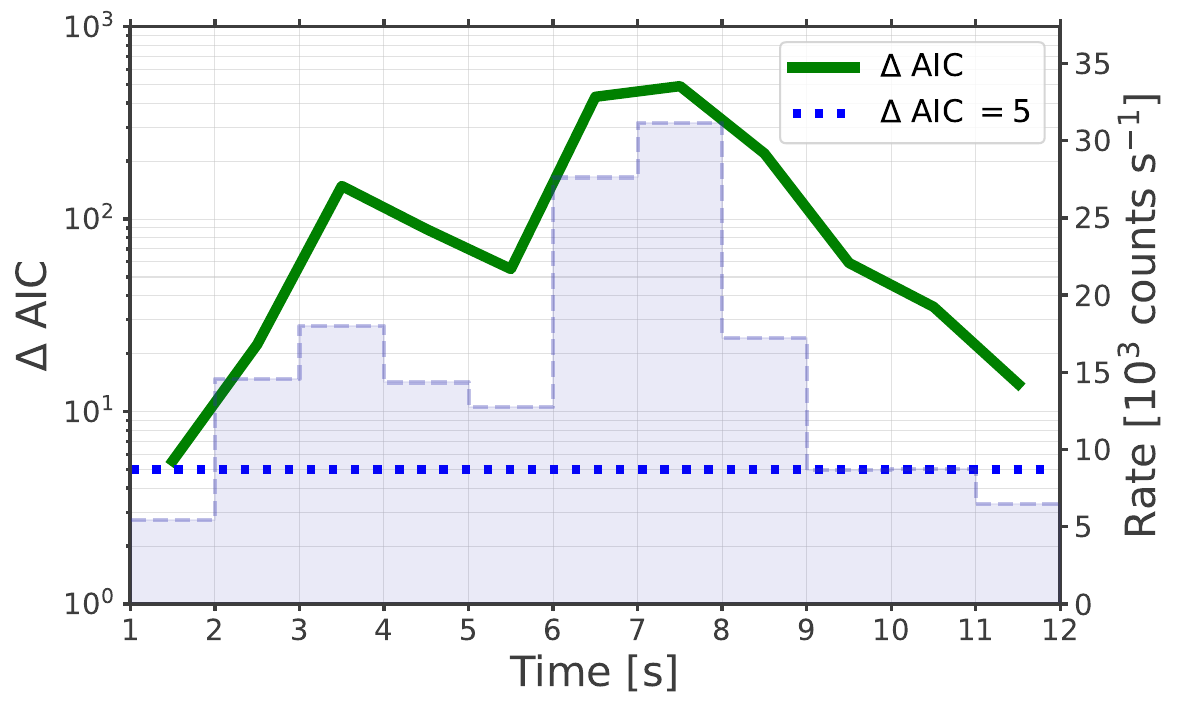}
    \caption{AIC difference between the KRA model and the Band function for every time bin. The dotted blue line shows the limiting value $\Delta {\rm AIC} = 5$, above which the KRA model is significantly preferred compared to the Band function. We see that in every time bin KRA is the preferred model. Also note the logarithmic y-axis. The light blue histogram is the binned light curve, with values on the y-axis to the right.}
    \label{fig:AIC}
\end{figure}

We compare the KRA model fits to Band function fits and the performance is evaluated by considering their respective fit statistic. We use the Akaike Information Criterion (AIC) \citep{Akaike74}, which accounts not only for the quality of the fit but also for the number of free model parameters. We define $\Delta {\rm{AIC}} = \rm{AIC}_{\rm{Band}} - \rm{AIC}_{\rm{KRA}}$ and use $\Delta{\rm{AIC}} \geq 5$ \citep[e.g.][]{Burnham04, Acuner20} as a limit where KRA is significantly preferred over the Band function. In Figure \ref{fig:AIC}, the $\Delta{\rm{AIC}}$ values from the fits are shown with the solid green line and the blue dotted line shows the limiting value of $\Delta{\rm{AIC}} = 5$. It is clear from the figure that the KRA is significantly preferred over the Band model fits for all time bins. The result is in agreement with previous works \citep{Gompertz23, Peng24}, which found that a model with a single low-energy power law (like Band) is not enough to capture the complexity of the spectral curvature below the peak in GRB 211211A. However, this is one of the first times a physical model performs better compared to the flexible and empirical Band function \citep[see also][]{Mei25}.

\section{Evolution of the sub-photospheric shock properties}\label{sec:results_RMS}

\begin{figure*}
    \centering
        \includegraphics[width = \columnwidth]{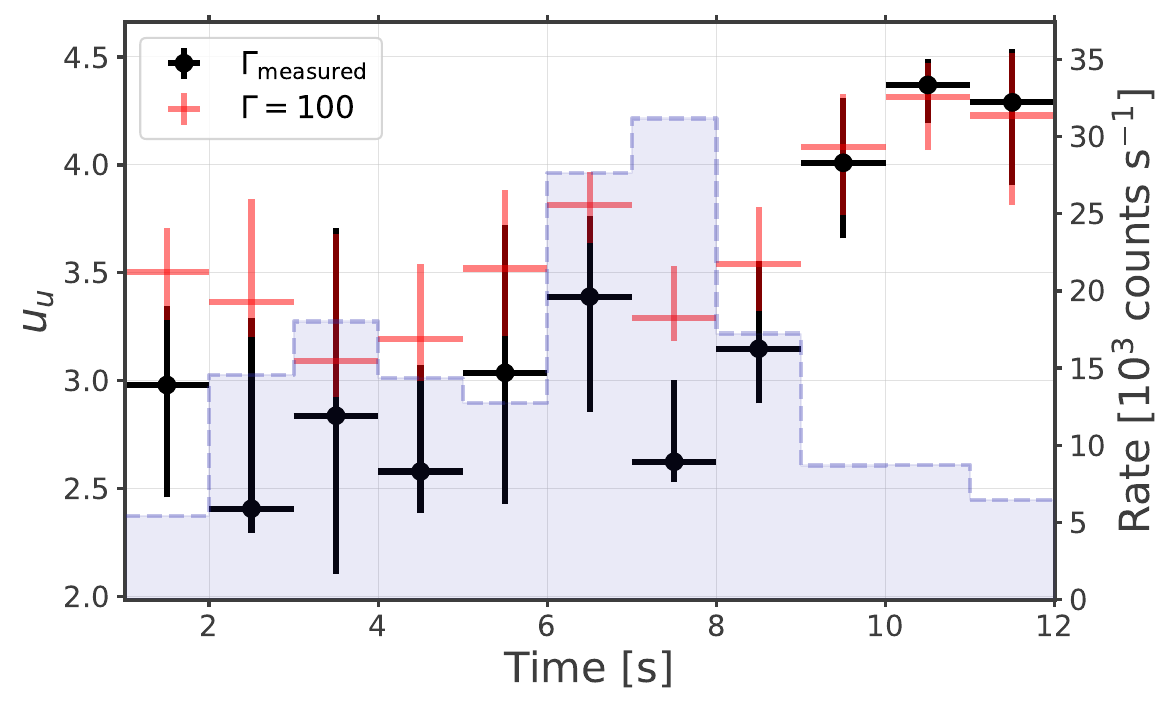}
        \includegraphics[width = \columnwidth]{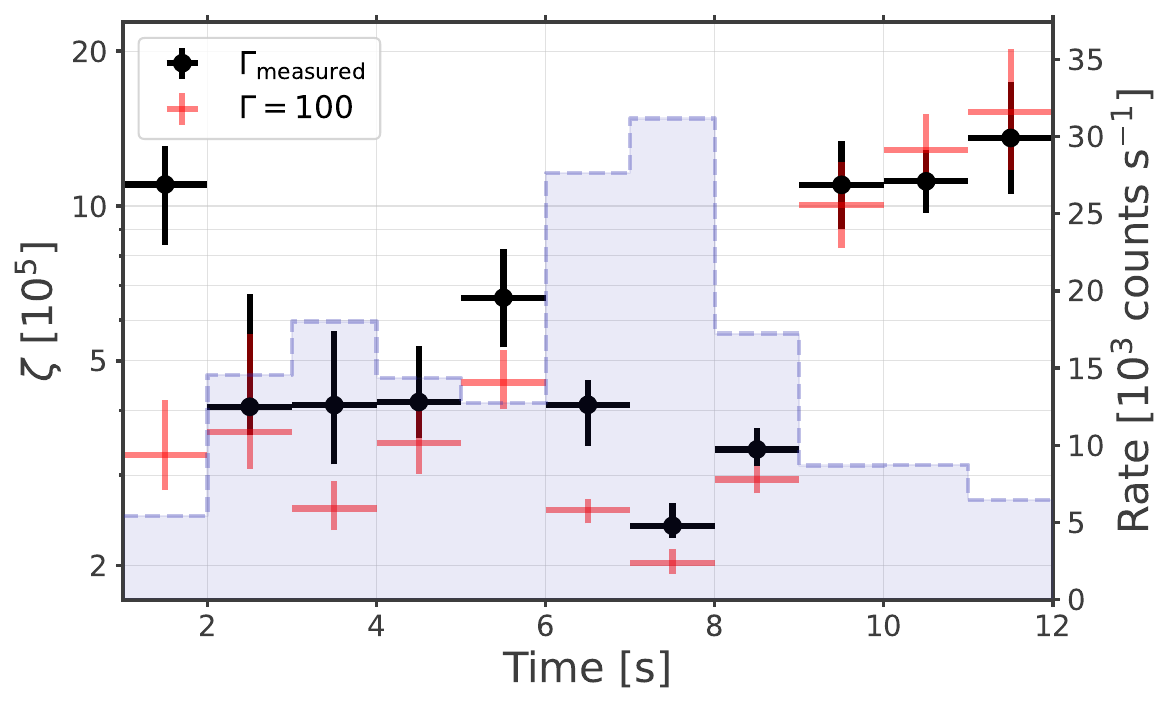}
        \includegraphics[width = \columnwidth]{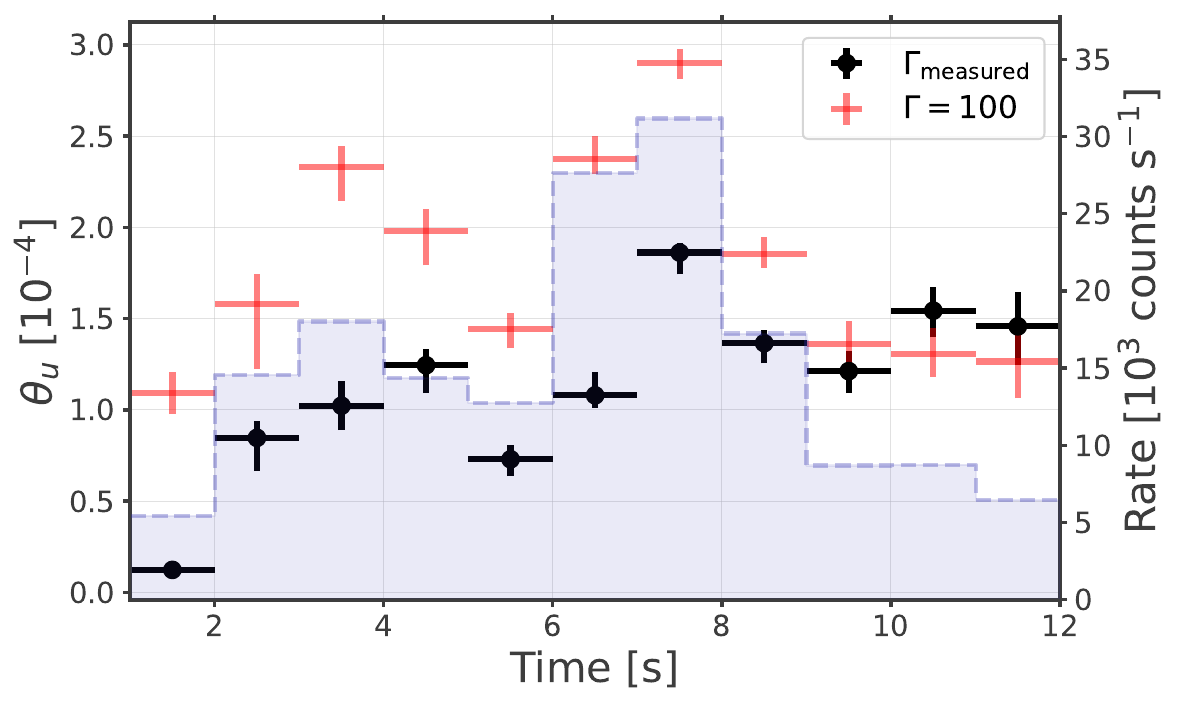}
        \includegraphics[width = \columnwidth]{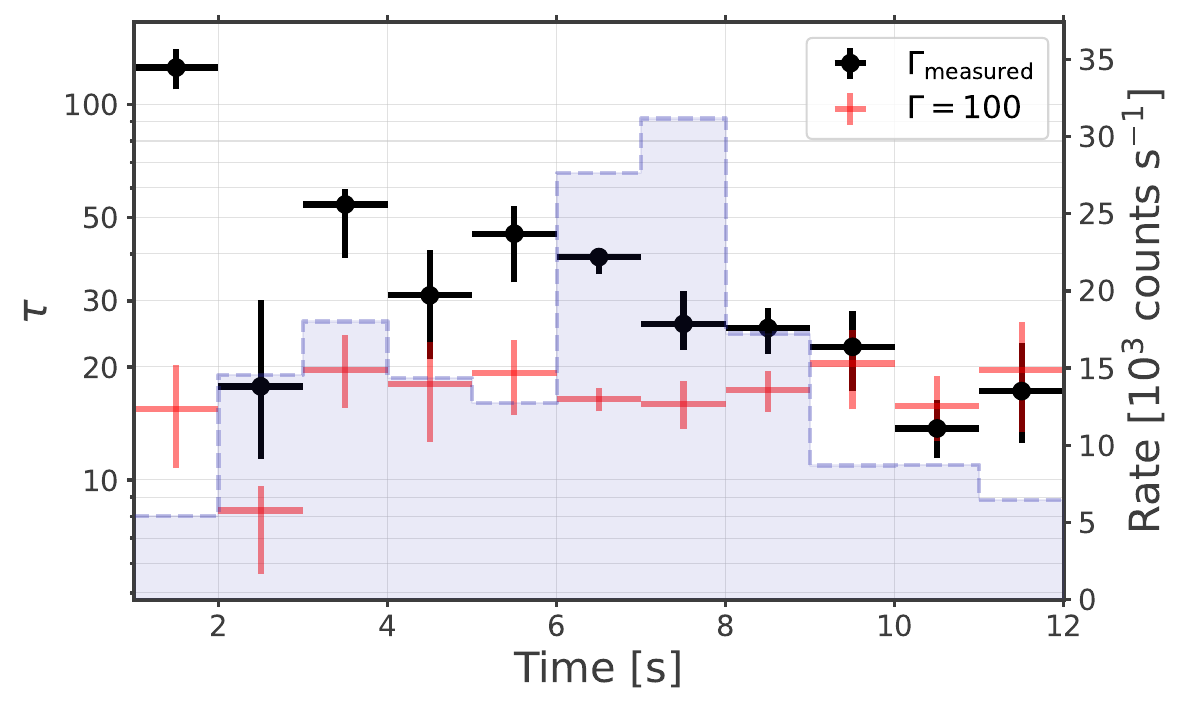}
    \caption{Time evolution of the RMS parameters from the Bayesian inference. The different colors represent different values for $\Gamma$, black is the measured $\Gamma$ from \S \ref{sec:gamma}, and red is a constant $\Gamma = 100$. The y-error bars are the 68\% confidence interval. Upper left is $u_u$, upper right is $\zeta$, bottom left is $\tu{}$, and bottom right is $\tau$. The light blue histogram is the binned light curve, with values on the y-axis to the right.}
    \label{fig:RMS_evolution}
\end{figure*}

The time-evolution of the RMS parameters is reported in Table \ref{tab:fit} and shown in Figure \ref{fig:RMS_evolution}. In the figure, the dimensionless specific momentum, $u_u$, is shown in the top left panel, the photon-to-proton ratio, $\zeta$, in the top right panel, the upstream temperature, $\tu{}$, in the bottom left panel, and the optical depth where the collision occurs, $\tau$, in the bottom right panel. The results from Table \ref{tab:fit} are shown with black markers, with the mean value and 68\% confidence intervals obtained by converting the full posterior distribution from \S \ref{sec:results_spectra} using the scheme outlined in \S \ref{subsec:convert}. For comparison, we also show the results from a similar analysis but for a constant Lorentz factor, $\Gamma = 100$, with red markers.

As pointed out in \S \ref{subsec:spectral_shape}, the last three time bins (9-12 s) have a different spectral shape. In Figure \ref{fig:RMS_evolution}, it can be seen that the corresponding RMS parameters are also noticeably different. Due to this difference, we treat the two cases separately, first focusing on the time-evolution during 1-9 seconds in \S \ref{subsec:1-9} and continue with the last three time bins in \S \ref{subsec:relativistic}.

\subsection{Interval 1-9 s: Strong, mildly relativistic shock}\label{subsec:1-9}

Here, we discuss each of the four RMS parameters in turn, focusing on the results obtained with the measured Lorentz factor from \S \ref{sec:gamma}. The dependence of the results on the Lorentz factor, i.e., a comparison between the black and red data points, is given in \S \ref{subsubsec:sensitivity}.

\subsubsection{\texorpdfstring{The dimensionless specific momentum, $u_u$}{The dimensionless specific momentum}}

The dimensionless specific momentum, $u_u = \gamma_u \beta_u$, is large ($>1$), which is necessary for the shock to be strong and dissipate a large amount of the kinetic energy \citep{Kobayashi97, Daigne98}. This, together with $\tau$ being quite low, indicates that the radiation efficiency is large. The efficiency is further analyzed in \S \ref{subsec:pair_efficiency}. 

The values of $u_u$ are mostly determined by the best-fit values of $\tr{}$, see Equation \eqref{eq:theta_r}. A complication in the determination of $\tr{}$ is the effect of thermalization on the spectrum. Photons that enter the downstream with comoving energy $\epsilon > 2 / \tau$ have time to down-scatter before they reach the photosphere. Therefore, when $4\tr{} = \epsmax{} > 2 /\tau$, all high-energy photons down-scatter to $2 / \tau$ and the maximum energy at the photosphere is determined solely by the optical depth and independent on the initial value of $\epsmax{}$. In this case, the peak energy can be used to get an estimate of the optical depth where the emission took place \citep{Alamaa24}. However, it becomes difficult to constrain $u_u$.

On the other hand, this complication is avoided when $2 / \tau$ is smaller than four times the downstream Compton temperature,\footnote{The Compton temperature is defined as the temperature the electrons obtain when in Compton equilibrium with the photons. Since the photons vastly outnumber the electrons, to a good approximation, the electrons are kept at the Compton temperature at all times.} $\tC{}$, since the photon cooling stops once $\epsilon \gtrsim 4\tC{}$. In the case, when $4\tC{} > 2 / \tau$, the observed peak energy is instead determined by the value of $\tC{}$, which in turn depends on $\tr{}$.

When $y_r > 1$, the Compton temperature in the immediate downstream is comparable to $\theta_r$ and, thus, $\tC{}$ is typically larger than $2 / \tau$ \citep[see Figure 2 in][]{Bagi25}. Therefore, we expect the value of $\theta_r$ to be constrained by the observed peak energy in the interval 1-9 s, where $y_r > 1$. Indeed, from Figure \ref{fig:corner_KRA}, it is clear that large values of $\theta_r$ are disfavored by the data. This shows that the intrinsic maximum energy can in some cases be deduced even after the occurrence of severe thermalization of high-energy photons between the dissipation site and the photosphere.

\subsubsection{\texorpdfstring{The photon-to-proton ratio, $\zeta$}{The photon-to-proton ratio}}

The photon-to-proton ratio, $\zeta = n_\gamma / n_p$, has values of a few $10^5$, except the first time bin which is slightly larger at $\sim 10^6$. These are around the expected values for RMSs in GRBs \citep{Bromberg11, Levinson12}. The figure shows that the ratio is anti-correlated to the photon count rate, which is closely related to $\Nob{}$. Fewer photons per proton means that a smaller number of photons share the same dissipated energy, which is initially carried in the form of kinetic energy by the upstream protons. Thus, a low $\zeta$ means that a larger fraction of photons reach higher energies, leading to a harder power-law slope between the two breaks \citep[e.g. Figure 2 in][]{Samuelsson22}. The anti-correlation, therefore, shows that the spectrum is harder when the photon flux is larger, in agreement with previous investigations \citep[the $\alpha$-intensity correlation, e.g., ][]{Crider97, Ryde19, wang23}.

\subsubsection{\texorpdfstring{The upstream temperature, $\tu{}$}{The upstream temperature}}

The upstream temperature, $\tu{}$, is $\sim 10^{-4}$ (in the comoving frame and in units of electron rest mass energy) during 2-9 seconds, while the first time bin is lower at $\sim 10^{-5}$. The errors for $\tu{}$ are quite small because it is constrained by the low-energy curvature, which is well recorded in the observed spectrum for all time bins (see Figure \ref{fig:spectral_evolution}).
Assuming a pure fireball scaling, the upstream temperature can be expressed as

\begin{equation}\label{eq:tu_disc}
    \tu{} = \theta_0 \lr{\frac{\rcoll{}}{r_0}}^{-2/3} \lr{\frac{\Gamma}{\Gamma_0}}^{-1/3}
\end{equation}

\noindent where $\rcoll{}$ is the collision radius, $r_0$ is the radius at the base of the jet where the acceleration starts, and $\theta_0$ and $\Gamma_0$ are the temperature and Lorentz factor at $r_0$, respectively. In this work, we find $\rcoll{} = \rph{} / \tau \lesssim 10^{12} \cm{}$, and using $\Gamma_0 \gtrsim 1$, we require $r_0 \sim 10^7 \cm{}$ and $\kB{} T_0 = m_e c^2 \theta_0 \sim 1 \MeV{}$ to be consistent with the measured values of $\tu{}$. Values of $r_0 \sim 10^7 \cm{}$ are consistent with other studies of GRBs \citep{Iyyani13, Larsson15}. However, larger values of $r_0 \sim 10^9 - 10^{10} \cm{}$ have been suggested both theoretically \citep{Thompson07}, and supported by numerical simulations \citep{Morsony07, Lazzati09, Gottlieb19}. Equation \eqref{eq:tu_disc} assumes a constant radial width of the relativistic wind, that is, the comoving density decreasing as $\rho \propto r^{-2}$. If the density decreases quicker as a function of $r$, the temperature drops more rapidly with radius as well. This would allow for larger values of $r_0$ and/or $\theta_0$ (see \S \ref{subsec:phys_scenario}).

In Figure \ref{fig:RMS_evolution} we find a correlation between $\tu{}$ and the photon count rate. The temperature $T_0$ can be expressed in terms of the central engine luminosity, $L$, as

\begin{equation}
    T_0 = \lr{\frac{L}{16 \pi \sigma r_0^2 \Gamma_0^2}}^{1/4}.
\end{equation}

\noindent Under the assumption that the photon count rate is correlated to the central engine luminosity, together with Equation \eqref{eq:tu_disc}, this correlation is expected.

The simple calculation above ignores the impact of the Lorentz factor on both $\tu{}$ and the photon count rate. Indeed, we find that the correlation between the photon count rate and the product $\tu{} \Gamma \propto \Tob{}$ (see Equation \eqref{eq:Tob}) is stronger than the correlation to $\tu{}$ alone. This is likely caused by a higher Lorentz factor increasing the Doppler boost, thus increasing $\Tob{}$, while simultaneously increasing the beaming, therefore increasing $\Nob{}$.

\subsubsection{\texorpdfstring{The optical depth, $\tau$}{The optical depth}}

For the optical depth, $\tau$, our results show that the shock occurs at a low to moderate optical depth ($\sim 20-50$) during 2-9 seconds and at a larger optical depth for the first time bin. Since the observed spectrum is broad it is still significantly altered by the shock, which implies a low/moderate optical depth. If the shock had occurred at larger $\tau$, the thermalization would be more efficient and the emitted spectrum would be closer to thermal. The first time bin has a larger optical depth, and thus one might expect a narrower spectrum. However, the width of the observed spectrum is still similarly broad to the other time bins. The reason is that the initial width in the immediate downstream is larger, due to a colder upstream temperature, i.e., a lower $\tu{}$. This illustrates that observed spectra can remain highly nonthermal, even when originating from optical depths of $\sim 100$.

To consistently find low to moderate optical depths could be explained by an observational bias. Shocks that occur closer to the photosphere are brighter, since the photons suffer less adiabatic cooling before decoupling at the photosphere. Therefore, if similar shocks occur at various optical depths in the outflow, the observed emission would be dominated by shocks from lower optical depths.

\subsubsection{\texorpdfstring{Sensitivity of the results on $\Gamma$}{Sensitivity of the results on the Lorentz factor}}\label{subsubsec:sensitivity}

The red markers in Figure \ref{fig:RMS_evolution} shows the time-evolution of the RMS parameters for a constant $\Gamma = 100$ (see also Figure \ref{fig:gammas}), which is lower than all the measured Lorentz factors during 1-9 seconds. From Figure \ref{fig:RMS_evolution} it is clear that the values of $u_u$ and $\tu{}$ increase and the values of $\zeta$ and $\tau$ decrease, when the Lorentz factor decreases. That $u_u$ and $\tu{}$ increase is expected since these two parameters regulate the comoving break energies and all comoving energies need to be higher for a lower Lorentz factor in order to match the observations. Similarly, the optical depth decreases to avoid a too large adiabatic degrading.

In order to explain the $\zeta$-dependence, we note that the kinetic energy density in the upstream is approximately $(\gamma_u~-~1)m_p c^2 n_p$, which is dissipated in the shock and converted into a downstream photon energy density $\epsd{} n_\gamma m_e c^2$. Solving for the photon-to-proton ratio gives

\begin{equation}
    \zeta \approx \frac{\gamma_u - 1}{\epsd{}} \frac{m_p}{m_e}.
\end{equation}

\noindent Similar to $u_u$ and $\tu{}$, the comoving energy $\epsd{}$ and temperature $\tr{}$ need to increase when $\Gamma$ decreases, in order to match the observations. For a mildly relativistic shock with $\beta_u \sim 1$, Equation \eqref{eq:theta_r} gives that $\gamma_u \propto \tr{}^{1/2}$. Thus, approximately $\zeta \propto \tr{}^{1/2}/\epsd{}$ and as both $\epsd{}$ and $\tr{}$ are inversely proportional to the Lorentz factor, $\zeta$ will decrease with decreasing $\Gamma$. Interestingly, since in the nonrelativistic limit, $\gamma_u - 1 = \beta_u^2 / 2$ and $\beta_u^2 \propto \tr{}$, the Lorentz-factor dependence of $\zeta$ vanishes. 

The correlation between $\tu{}$ and $\Nob{}$ mentioned previously is even stronger when the Lorentz factor is constant. This is because the product $\tu{} \Gamma \propto \Tob{}$ is still correlated to $\Nob{}$ and thus with a constant Lorentz factor the correlation of $\tu{}$ has to be larger. This shows that assuming a constant value of the Lorentz factor over the whole burst duration can lead to artificial correlations.

\subsection{Interval 9-12 s: A potentially relativistic shock}\label{subsec:relativistic}

The KRA model is applicable to nonrelativistic and mildly relativistic shocks. In the KRA model this requirement corresponds to that the maximum energy of a photon is $\epsmax{} = 4\theta_r < 1$, i.e., below the electron rest mass energy. For relativistic shocks, other spectral shapes are expected due to Klein-Nishina effects, $\gamma\gamma$-pair production, and anisotropies \citep{Ito18, Lundman18}. Also, since $\langle \Delta \epsilon / \epsilon \rangle > 1$ the relative energy gain per scattering becomes large and a smooth power-law is no longer expected.

In our analysis, the last three time bins (9-12 s) have a reported value for $\theta_r$ in Table \ref{tab:fit} of $\geq 0.20$. Investigating the posterior distribution we see that the best-fit value is affected by our prior of $\theta_r < 0.25$, set by the nonrelativistic limit. Therefore, the derived RMS parameters are affected by our choice of priors and it is possible that the shock at this time is relativistic and consequently our model is not directly applicable. Thus, the obtained parameters should be treated with caution. 

However, the data indicates smooth power-law like spectra at all times, including the three last time bins, see Figure \ref{fig:RMS_evolution}. In the immediate downstream of relativistic RMSs, one expects more complex spectral features due to the complexities mentioned above. On the other hand, such features might be removed by thermalization before the photosphere and be obscured by geometrical broadening. This suggests that further investigations are needed of the spectral shape that is formed by increasingly relativistic RMSs. Moreover, if taken at face value, the RMS parameters for the last three time bins are quite similar to those during 1-9 seconds and most of the discussion in \S \ref{subsec:1-9} could still be applied. 

For the last three time bins, the measured Lorentz factors are very close to the constant Lorentz factor ($\Gamma = 100$). Thus, the difference in values between the black and red markers in Figure \ref{fig:RMS_evolution} is small.

\section{Discussion}\label{sec:discussion}

\subsection{Pair loading and dissipation efficiency}\label{subsec:pair_efficiency}

In \S \ref{sec:gamma}, we determined the Lorentz factor based on the observed variability time by using Equation (\ref{eq:gamma_tvar}). The Lorentz factor can also be determined based on the observed flux by using the expression $\rph{} = L\sigma_{\rm T}\kappa_\pm/4\pi m_p c^3 \Gamma^3$, together with $L = 4\pi d_L^2 \Fob{}/\eff{}$, where $\sigma_{\rm T}$ is the Thomson cross section, $\Fob{}$ is the observed energy flux, $\kappa_\pm = (n_e + n_\pm)/n_e$ is the pair multiplicity of the outflow, and $\eff{}$ is the radiation efficiency \citep{Peer07, Wistemar25}. Using the above equations, together with Equations \eqref{eq:gamma_tvar} and \eqref{eq:r_ph}, one can solve for the unknown ratio of $\kappa_\pm / \eff{}$, arriving at

\begin{equation}\label{eq:kappa_eff}
    \frac{\kappa_\pm}{\eff{}} = \lr{\frac{5.4 k_B m_p^2 c^6}{\sigma_T^2 \sigma} \frac{\Nob{} \Gamma^8}{d_L^{2} (1 + z)^{4} \Tob{}^{3} \Fob{}^{2}}}^{1/2}.
\end{equation}

\begin{figure}
    \centering
    \includegraphics[width = \columnwidth]{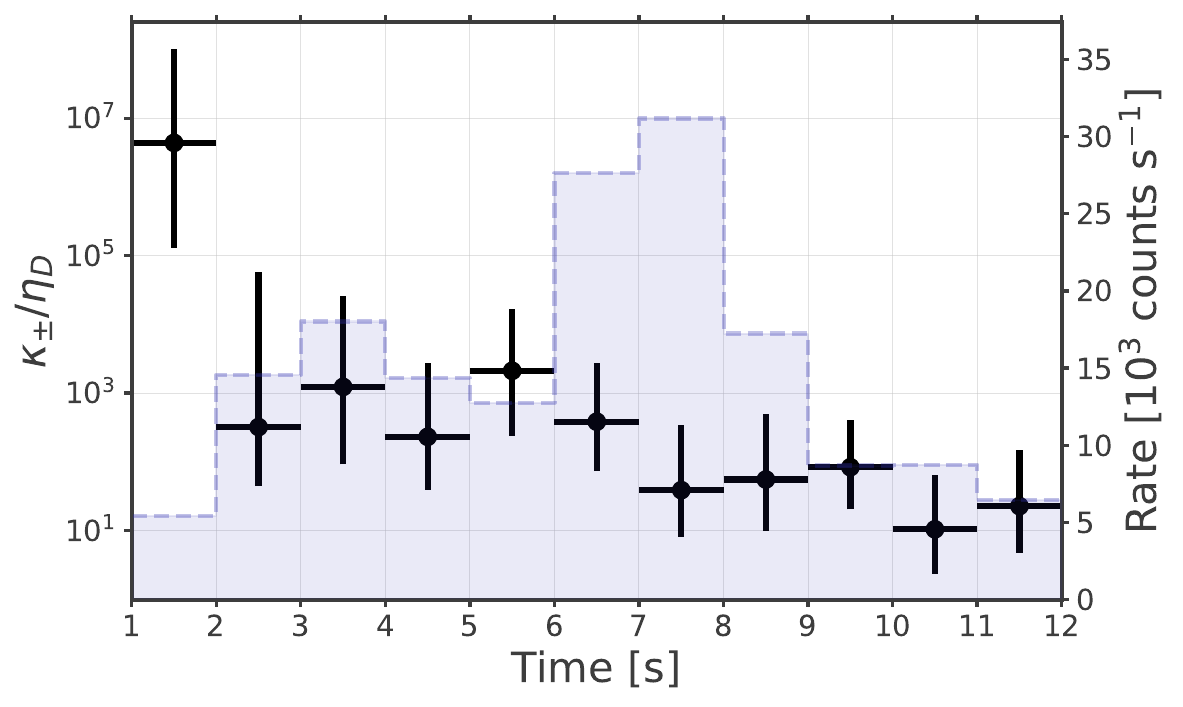}
    \caption{Time evolution of the ratio of the pair loading, $\kappa_\pm$, and radiation efficiency, $\eff{}$. Equation \eqref{eq:kappa_eff} is used to calculate the values and the 68\% confidence intervals. The light blue histogram is the binned light curve, with values on the y-axis to the right.}
    \label{fig:kappa_eff}
\end{figure}

\noindent To estimate the ratio, we use the values for $\Gamma$, $\Nob{}$, and $\Tob{}$ from Table \ref{tab:gamma}, $z = 0.076$ \citep[][]{redshift21}, and the value of $\Fob{}$ from integrating the fitted KRA spectrum (similarly to $\Nob{}$). The estimations of the ratio, ${\kappa_\pm}/{\eff{}}$, are shown in Figure \ref{fig:kappa_eff}, together with the 68 \% confidence intervals. 

Equation \eqref{eq:kappa_eff} shows that the ratio has a strong dependence on the Lorentz factor, as $\kappa_\pm / \eff{} \propto \Gamma^4$. Inserting $\Gamma$ from Equation \eqref{eq:gamma_tvar}, one gets $\kappa_\pm / \eff{} \propto \tvar{}^{-4} \Tob{}^{-15/2}$. Thus, we note that the parameter dependence, specifically on the temperature, is very strong. The ratio shown in Figure \ref{fig:kappa_eff} is calculated using the variability time from \citet{Veres23}, but other works using different methods have found larger variability times, which would greatly lower the ratio. Furthermore, an error in the estimation of $\Tob{}$ by a factor of two would alter the ratio by roughly two orders of magnitude. 

However, if we take the values in Figure \ref{fig:kappa_eff} at face value, the values of the ratio is $\sim 10^2-10^3$ during the most luminous part of the burst (2-9 s). Since we found large values for $u_u$ and low optical depths $\tau$, the radiation efficiency should be high. This is difficult to reconcile if $\kappa_\pm = 1$, as this would imply a low efficiency. Therefore, the expected high efficiency points towards a pair-loaded jet.

One possible origin of such a pair loading is that the jet of GRB 211211A needs to propagate through a neutron-rich wind from the NS-NS or NS-BH merger \citep{Rastinejad22, Troja22}. The source of pairs could then come from the sideways diffusion of neutrons from the neutron-rich wind into the jet \citep{Meszaros00, Levinson03, Preau21}. The newly diffused neutrons, together with baryons in the jet, produce pions from inelastic collisions. The pions would decay into high energy photons that pair produce and induce an electromagnetic cascade. Therefore, the jet could be loaded with a large number of electron-positron pairs. The pair loading is determined by the e$^{\pm}$ injection rate following the cascade, and e$^{\pm}$ annihilation rate \citep{Derishev99, Beloborodov10, Murase13}.

The KRA model does not account for any radiation from pairs, and it is therefore somewhat inconsistent to infer a pair loading from a KRA fit. However, given that the photon-to-proton ratio, $\zeta$, is around a few $10^5$ (see Table \ref{tab:fit}), the photons still vastly outnumber the number of pairs for typical efficiencies of $\eff{} \sim 0.1$. Thus, it is plausible that the pairs do not have a strong impact on the observed spectrum. One effect the pairs may have is the creation of a high-energy power law from cascades \citep{Beloborodov10}.

The first time bin also behaves differently here, with $\kappa_\pm / \eff{} \sim 10^7$, much larger than the other time bins. This occurs because of the strong dependence on $\Gamma$ in Equation \eqref{eq:kappa_eff}, together with the larger Lorentz factor for the first time bin. Typically $\kappa_\pm$ is assumed to be smaller than $m_p / m_e = 1836$ for the inertia of the flow to be dominated by the protons. To satisfy this limit, $\eff{}$ would need to be exceptionally low ($\lesssim 4 \times 10^{-4}$).

\subsection{Division between main and extended emission}\label{subsec:last_bins}

As previously mentioned, all time bins are well fitted by the KRA. This includes the last three time bins, even though they stand out both in shape and derived parameter values, see \S \ref{subsec:relativistic}. Interestingly we note that the spectral shape of the last three time bins is remarkably similar to the spectral shape of the extended emission between 17$-$25 s. The similar spectral shape could point to the same emission process for the different times. Potentially then, the extended emission in GRB 211211A already starts at $\sim 9$ seconds after the trigger.

We also note the similar light curve morphology at $9-12$ s and $17-20$ s, as evident in Figure \ref{fig:light_curve_full}. This, together with the similar spectral shape, could be used to argue that the intervals are from the same emission episode \citep{Hakkila21}.

\subsection{Physical scenario in GRB 211211A with different emission models}\label{subsec:phys_scenario}

Both a photospheric (as demonstrated here) and a synchrotron framework \citep[][]{Gompertz23, Mei25} can fit the prompt observations of GRB 211211A significantly better than the Band function. The reason for this is that the data is not well described by a single power law at low energies. Some type of additional curvature is needed and this is something that a marginally-fast cooling synchrotron spectrum and our RMS model can provide. However, 
the physical scenarios that they would occur in are fundamentally different. Moreover, both models have some limitations in terms of the plausibility and consistency of the inferred physical conditions.

In a fast cooling synchrotron framework, $\gamma_m$ is the injection Lorentz factor and $\gamma_c$ is the cooling Lorentz factor, responsible for the peak energy, $E_p$, and the break energy, $E_c$, in the observed spectrum, respectively. Therefore, when both $E_p$ and $E_c$ can be inferred from the data, the ratio $\gamma_m / \gamma_c = \lr{E_p / E_c}^{1/2}$ is known \citep[since $E \propto \gamma^2$;][]{Rybicki_Lightman79}. Using the best-fit values for $E_p$ and $E_c$ as found by \citet{Mei25} gives an average value for this ratio as $\gamma_m / \gamma_c \sim 4$ during the main emission in GRB 211211A. Thus, the emitting particles were marginally fast cooling \citep[$\gamma_m \gtrsim \gamma_c$,][]{Daigne11}.

The closeness of the two characteristic electron Lorentz factors, $\gamma_m$ and $\gamma_c$, in GRB 211211A is consistent with the findings from synchrotron fits in other GRBs \citep{Burgess20}. Since there is no a priori reason for why this would be, this presents a fine-tuning problem that is, arguably, one of the major theoretical challenges with contemporary synchrotron emission models \citep[see][for potential solutions]{Ghisellini20, Daigne25}. We note that the possible effect of inverse Compton cooling, specifically in the Klein-Nishina regime \citep{Derishev01, Nakar09, Bosnjak09, Geng18}, was neglected in \citet{Mei25}. Depending on the strength of the inverse Compton cooling, it can help to generate a spectrum with a low-energy break below the peak. Other ways to alleviate the problem is to allow for internal relativistic motion \citep[so called mini-jets; ][]{Beniamini18, Burgess20}, or emission from both forward and reverse shocks \citep{Daigne98, Rahman24}.

An important quantity in synchrotron emission is the magnetic field, and its comoving field strength at the emission site can be solved for as \citep[see e.g. Equation (2) in][]{Begue22}

\begin{equation}
    B = \lrb{\frac{48 \pi m c^3 h e}{\sigma_T^2} \lr{\frac{m}{m_e}}^{4} \frac{\Gamma^3}{E_c r_e^2}}^{1/3} \approx 25 \; \frac{\Gamma_2}{r_{e, 14}^{2/3}} \; \rm{G}
\end{equation}

\noindent where $m$ is the mass of the emitting particle, $h$ is the Planck constant, $e$ is the elementary charge, $r_e$ is the emitting radius, and we have used the notation $Q_x = Q / 10^x$. For the numerical value, we used $E_c = 100 \keV{}$ as measured by \citet{Mei25} and assumed that the emitting particles are electrons, which results in a magnetic field strength that is much lower than expected \citep{Ghisellini20}. One way to increase the magnetic field is to instead consider protons as the radiating particles \citep{Ghisellini20}, however this requires a very large magnetic luminosity \citep{Florou21, Begue22}. 

The photospheric framework investigated here also presents challenges, both in terms of model assumptions and with regards to the physical interpretation. The potentially relativistic shock in the final time bins, see \S \ref{subsec:last_bins}, limits the applicability of the KRA model. In the case of a relativistic shock, core assumptions are not valid, and while the fits remain good, the inferred physical conditions may be incorrect. Additionally, the model exhibits parameter degeneracies in the absence of a constrained Lorentz factor. Although this is addressed in \S \ref{sec:gamma} using the method in \citet{Wistemar25}, the inferred results are sensitive to the value of the variability timescale, which introduces systematic uncertainty.

With regards to the physical scenario, we consider the high shock speeds and the cool upstream temperatures to be the most surprising. In an internal collision framework, mildly relativistic relative motion ($u_u \sim 3$) is quite difficult to achieve \citep{Kobayashi97, Daigne98, Levinson20, Samuelsson22}. Observational bias could play a role here, since the radiation efficiency increases with increasing $u_u$.  The low upstream temperatures found seem to be in tension with numerical simulations, which suggest strong collimation up to large distances \citep{Morsony07, Lazzati09, Gottlieb19}, effectively increasing the value of $r_0$ to $\sim 10^{10}~$cm. Even with $\rho \propto r^{-3}$, this is incompatible with the values of $T_{\rm obs}\sim 5~$keV, $r_{\rm ph}\sim2\times 10^{13}~$cm, and $\Gamma \sim 300$ found in this work, which gives a maximum value of $r_0 \sim {\rm few} \ 10^{8}~$cm for $k_{\rm B}T_0 \sim 1~$MeV. Whether the parameter values found for GRB 211211A are typical to the whole GRB population or not remains to be seen.

Another difference between the synchrotron and the photospheric emission models, tied not to the physical scenario but to the observations, comes from the broad-band spectral shape. Beyond the typical gamma-ray band of around $10\keV{} - 10 \MeV{}$, spectra necessarily differ, with RMS spectra being much narrower compared to synchrotron spectra. Unfortunately, broad-band spectra are typically not available for the prompt phase, which makes this measurement difficult.

Occasionally, simultaneous observations in gamma-rays and optical are available. For instance, \citet{Oganesyan19} showed a clear consistency with synchrotron over 5 orders of magnitude in energy. In GRB 211211A there are no simultaneous observations during the main emission (1-12 s). However, there is an optical data point at 163 s \citep[see Figure 6 in][]{Gompertz23}, which is significantly lower than the extrapolation of the synchrotron spectral slope with a photon index of $-2/3$, requiring an additional break due to synchrotron self-absorption.

Finally, we note that having two different physical models, synchrotron \citep[e.g.,][]{Mei25} and photospheric, allows for a statistical model comparison \citep[see for instance][]{Meng18}. Such a model comparison could be an interesting future study.
\section {Conclusions}\label{sec:conclusion}

The main emission phase in GRB 211211A has broad and nonthermal spectra, including a strong curvature below the peak around a second spectral break. It has been interpreted in a synchrotron framework \citep{Gompertz23, Mei25} and with a nonthermal component with an additional subdominant BB \citep{Peng24}. In this paper, a photospheric model based on RMSs dissipating energy below the photosphere \citep[KRA;][]{Samuelsson22} has been used to fit the main emission of GRB 211211A. We find that the observed spectra are well fitted by the KRA model, and the evolution of the spectral shape is shown in Figure \ref{fig:spectral_evolution}. This result shows that dissipative photospheric emission can be very broad. The KRA model fits are significantly better than the commonly used empirical Band function in all time bins studied, as shown in Figure \ref{fig:AIC}. From the KRA model fits, the RMS properties can be derived and their time evolution is shown in Figure \ref{fig:RMS_evolution}. We find a strong shock, with $\gamma_u \beta_u \sim 3$, occurring at a moderate optical depth of $\tau \sim 35$. The upstream is relatively cold, with $\theta_u = k_{\rm B} T_u / m_e c^2 \sim 10^{-4}$, while the value for the photon-to-proton ratio, $\zeta \gtrsim 10^5$, is similar to theoretical expectations. These numbers are sensitive to the value of the bulk Lorentz factor, $\Gamma$, which was measured following the method of \citet{Wistemar25}, giving rise to the time-resolved measurement of $\Gamma$ shown in Figure \ref{fig:gammas}. However, the obtained RMS parameter values do not depend strongly on the Lorentz factor, as evident by comparing the red and black markers in Figure \ref{fig:RMS_evolution}. With the Lorentz factor determined, other burst properties can be estimated, such as the ratio of the jet pair loading and the radiation efficiency, $\kappa_\pm / \eff{}$. As shown in Figure \ref{fig:kappa_eff}, we find indications that the jet is pair loaded, although the ratio is highly sensitive to the parameters (see Equation \eqref{eq:kappa_eff}).

Our results thus show that sub-photospheric RMSs can provide a physical explanation for the prompt emission features of GRB 211211A, challenging the assumption that such spectral shapes necessarily originate from optically-thin emission mechanisms \citep[e.g., ][]{Lloyd2000, Iyyani13, Burgess20, Ryde22}. The results also highlight that the MeV-band spectral shape alone cannot uniquely determine the emission mechanism. Instead, the physical requirements of the alternative scenarios need to be compared and assessed. Further resolution can also come from simultaneous, broadband observations.

\section*{Acknowledgments}
    We thank Dr. Peter Veres for useful discussions and the anonymous referees for insightful comments. F.A. is supported by the Swedish Research Council (Vetenskapsrådet, 2022-00347). F.R. acknowledges support from the Swedish National Space Agency (2021-00180 and 2022-00205). This research has made use of data and/or software provided by the High Energy Astrophysics Science Archive Research Center (HEASARC), which is a service of the Astrophysics Science Division at NASA/GSFC.

\section*{Data availability}

The data from the \textit{Fermi} team used in the fitting procedure is provided by HEASARC, publicly available at \url{https://heasarc.gsfc.nasa.gov/}. The data underlying the results in this article will be shared on reasonable request to the corresponding author.

\bibliographystyle{mnras}
\bibliography{references}

\label{lastpage}
\end{document}